\documentclass{emulateapj-rtx4}
\usepackage{natbib}
\usepackage{epsf,graphics}
\usepackage{subfigure,graphicx}
\usepackage{bm}
\usepackage{amssymb}
\usepackage{amsbsy}
\usepackage{amsmath}
\usepackage{gensymb}
\usepackage{color}

\begin{document}

\title{Inferring the concentration of dark matter subhalos perturbing strongly lensed images}

\author{Quinn E. Minor}
\affiliation{Department of Science, Borough of Manhattan Community College, 
City University of New York, New York, NY 10007, USA}
\affiliation{Department of Astrophysics, American Museum of Natural History, 
New York, NY 10024, USA}
\author{Manoj Kaplinghat}
\affiliation{Department of Physics and Astronomy, University of California, 
Irvine CA 92697, USA}
\author{Tony H. Chan}
\affiliation{Department of Physics, City College of New York, New York, NY 10031, USA}
\author{Emily Simon}
\affiliation{Department of Astronomy, Columbia University, New York, NY 10027, USA}

\begin{abstract}

We demonstrate that the perturbations of strongly lensed images by low-mass dark matter subhalos are significantly impacted by the concentration of the perturbing subhalo. For subhalo concentrations expected in $\Lambda$CDM, significant constraints on the concentration can be obtained at HST resolution for subhalos with masses larger than about $10^{10}M_\odot$. Constraints are also possible for lower mass subhalos, if their concentrations are higher than the expected scatter in CDM. We also find that the concentration of lower mass perturbers down to $\sim 10^8M_\odot$ can be well-constrained with a resolution of $\sim 0.01''$, which is achievable with long-baseline interferometry. Subhalo concentration also plays a critical role in the detectability of a perturbation, such that only high concentration perturbers with mass $\lesssim 10^9M_\odot$ are likely to be detected at HST resolution. If scatter in the $\Lambda$CDM mass-concentration relation is not accounted for during lens modeling, the inferred subhalo mass can be biased by up to a factor of 3(6) for subhalos of mass $10^9 M_\odot$($10^{10} M_\odot$); this bias can be eliminated if one varies both mass and concentration during lens fitting. Alternatively, one may robustly infer the projected mass within the subhalo's perturbation radius, defined by its distance to the critical curve of the lens being perturbed. With a sufficient number of detections, these strategies will make it possible to constrain the halo mass-concentration relation at low masses in addition to the mass function, offering a probe of dark matter physics as well as the small-scale primordial power spectrum.
\end{abstract}

\keywords{gravitational lensing: strong -- dark matter -- galaxies: dwarf\vspace{1.0mm}}

\section{Introduction}\label{sec:intro}

A key prediction of the Cold Dark Matter (CDM) paradigm is the existence of a 
large number of dark matter subhalos around galaxies, with masses that go down 
to $< 1 M_\odot$ for the smallest halos \citep{green2004}. The vast majority of 
these subhalos are expected to be entirely devoid of stars, as their primordial 
gas would have been heated sufficiently by the ultraviolet background during 
reionization to escape the shallow potential wells of these subhalos. In the 
past few years, hydrodynamical cosmological simulations that include radiative 
transfer effects have estimated that star formation is suppressed entirely in 
dark matter halos with virial masses lower than $10^8$-$10^9 M_\odot$ 
\citep{sawala2016}, where the exact threshold depends on 
the details of the gas heating and cooling. The observed ultrafaint satellites 
of the Milky Way galaxy, such as those recently discovered in the Dark Energy 
Survey \citep{bechtol2015}, are thus expected to inhabit dark matter halos that 
lie just above this threshold \citep{brown2014}.  Detecting a large population 
of dark matter halos with masses $\lesssim 10^9 M_\odot$ is thus a crucial test 
of CDM.

Many well-motivated particle models for dark matter diverge from the CDM 
paradigm, with very different predictions for small-scale structure. For 
example, in warm dark matter models (WDM), the thermal motion of free-streaming 
dark matter particles erased small-scale structure before nonlinear collapse 
would have occurred; as a result, the number of dark matter subhalos is 
suppressed below a particular mass threshold (typically below $\sim 10^9 M_\odot$ for viable models; 
\citealt{bose2017}), and galaxies form with lower central densities 
\citep{lovell2014}.  In hidden sector models, dark matter particles 
can interact among themselves via a hidden force \citep{feng2009}.  A sufficiently high 
self-interaction cross section can lower the central density of low-mass halos and subhalos compared to that of 
CDM \citep{rocha2013,elbert2014}. However, the particle model of dark 
matter is not the only possible factor that can alter the small-scale matter 
power spectrum: certain types of inflation models can produce a running 
(departure from power law) in the primordial power spectrum \citep{kobayashi2011,ashoorioon2006,minor2015}, suppressing power at small scales 
 \citealt{garrison2014}). Whether primordial or through 
late-time effects, all of these alternative cosmological paradigms make 
distinct predictions for both the abundance and central densities of dark 
matter halos of a given mass.


Strong gravitational lensing provides a powerful probe of the dark matter 
distribution on dwarf galaxy scales, since the lensed images of a background source 
can undergo visible perturbations due to the presence of dark matter subhalos 
as well as dark matter halos that lie along the line of sight to the lens 
galaxy. Recently a few detections of dark substructures in gravitational lenses 
have been reported by observing their effect on highly magnified images.  Two 
of these were discovered in the SLACS dataset \citep{vegetti2010,vegetti2012}, 
while more recently, a subhalo was reported by \cite{hezaveh2016} to have been 
detected in an ALMA image of the lens system SDP.81 \citep{alma2015}. In order 
to test the expected mass function of halos in CDM, a robust estimate of the 
mass of these perturbing dark matter halos is very important; unfortunately 
however, the total inferred mass of the perturbers is highly dependent on their 
assumed density profile and tidal radius \citep{minor2017,vegetti2014b}. One approach to 
test CDM is to simply assume that the perturbers follow a Navarro-Frenk-White 
(NFW) profile (or a truncated form thereof), since the dark matter halos in CDM  N-body simulations are well fit by this profile \citep{navarro1996}. This approach is supported by recent results from hydrodynamical simulations that subhalos below $10^9M_\odot$ are near the threshold for star formation and hence may not be significantly altered by baryon physics \citep{fitts2017,sawala2016,ocvirk2016}. Even under this assumption, however, there is some expected scatter in the central density of halos of a given mass, embodied by the concentration parameter $c_{200} = r_{200}/r_s$; hence, assuming that subhalos follow a tight mass-concentration relation without scatter may also bias the inferred masses.

Estimating the mass of a perturbing subhalo is not the only quantity of 
importance, however. We argue that constraining the concentration of perturbing 
dark matter subhalos is equally important as the total mass, since different 
particle models for dark matter (or even inflation physics) can differ widely in their predictions for the central densities of dark matter halos \citep{lovell2014,rocha2013}. Thus, an important question is whether the 
concentration of perturbing dark matter halos can be meaningfully constrained  in addition to their masses. If the perturber's concentration  significantly affects the strength of the perturbation to the local lensing  deflection, then in principle meaningful constraints on the concentration are possible. Indeed, recently \cite{gilman2020} have shown that constraints on subhalo concentrations are possible for lensed quasars, at least in a statistical sense. If concentrations of individual subhalos can be constrained from their perturbations of lensed arcs, this approach may not only achieve a more  robust mass estimate, but may provide an additional probe of the small-scale power spectrum beyond simply testing the CDM halo mass function.

In this paper we simulate and model hundreds of mock lensing data to 
investigate the effect of dark matter subhalo concentration on lensing 
perturbations. We show that constraints on dark matter subhalo 
concentrations are indeed possible, and provide a powerful probe of the small-scale 
matter power spectrum. In addition, we will show that making the approximation that subhalos follow an exact 
mass-concentration relation (without scatter) during modeling of substructure perturbations can 
result in biased mass inferences, by up to a factor of $\sim$6 or so. Hence, even if 
dark matter is indeed cold and collisionless, scatter in the mass-concentration 
relation cannot be ignored if one aims to test the 
expected CDM halo mass function at dwarf galaxy scales via strong lensing.

We organize the paper as follows: in Section \ref{sec:mockdata}, we describe our lens modeling software and 
simulated data. In  Section \ref{sec:results}  we investigate how well the perturber's 
concentration can be constrained, both at HST resolutions as well as at higher 
resolutions expected from next-generation telescopes. In Section \ref{sec:detections}, we will investigate the effect of 
halo concentration on the detectability of substructure perturbations. In Section \ref{sec:mc_degeneracy} we will discuss the physical interpretation of the degeneracy between mass and concentration in terms of the size of the subhalo's lensing perturbation. Next, in Section 
\ref{sec:mbias} we explore the expected bias (in CDM) in the inferred perturber's mass 
if scatter in halo concentration is unaccounted for, while in Section 
\ref{sec:robust_mass} we demonstrate that bias in these cases is eliminated by 
using the mass estimator of \cite{minor2017}. Finally, in Section 
\ref{sec:c_limits} we discuss the physical interpretation of the concentration constraints, and in Section \ref{sec:tidal_truncation} we discuss the effect of tidal stripping on our results. We conclude in Section \ref{sec:conclusions}.

\section{Lens modeling and mock data}\label{sec:mockdata}

\subsection{Lens simulation and modeling procedure}\label{sec:lensmodel}

To investigate the effect of subhalo concentration, we generate a grid of mock 
data that explores a variety of different subhalo masses, concentrations, 
positions in the lens plane, and lens redshifts. For the primary lens we use a 
power-law ellipsoidal projected density profile, plus an external shear term in 
the potential. We adopt parameters such that it closely resembles the lens 
SDSSJ0946+1006, for which \cite{vegetti2010} detected a perturbing subhalo. For 
our primary mock data grid, we generate simulated images using the pixel size 
of the Hubble Space Telescope (HST) ACS camera ($0.045''$) and the actual HST point-spread function 
(PSF) pixel map used by \cite{vegetti2010}. For a small subset of these 
data, we will also generate and model a high-resolution version with $0.01''$ 
pixel size (described in Section \ref{sec:mc_constraints_hires}). The source galaxy 
is created using an elliptical Gaussian profile with two possible sizes, 
corresponding to unlensed widths of 0.03 and 0.06 arcseconds.

\begin{figure*}
	\centering
	\subfigure[mock source]
	{
		\includegraphics[height=0.36\hsize,width=0.41\hsize]{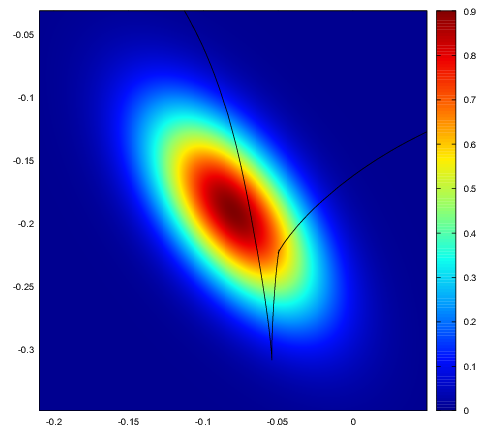}
		\label{fig:mocksrc}
	}
	\subfigure[mock data image]
	{
		\includegraphics[height=0.36\hsize,width=0.41\hsize]{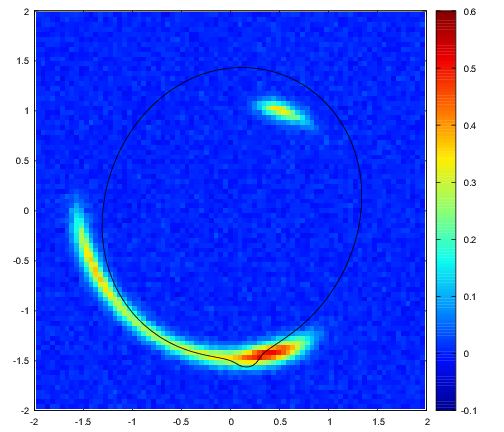}
		\label{fig:mockimg}
	}
	\caption{Mock source and data image, with a $10^{10}M_\odot$ perturber. In panel (a) the mock source is shown with caustic overlaid; note the kink in the caustic generated by the perturbing subhalo. In panel (b) the mock image is shown with critical curve overlaid, where a perturbation is evident in the lower arc.}
\label{fig:mockdata}
\end{figure*}

In our primary mock data images, the subhalo is modeled using a spherical 
Navarro-Frenk-White (NFW) profile \citep{navarro1996}, without any tidal truncation. As we show in 
Section \ref{sec:tidal_truncation}, for the majority of subhalo perturbations 
we expect the subhalo's tidal radius to be much larger than its NFW 
scale radius, since the most subhalos will be several tens to hundreds 
of kiloparsecs away from the host galaxy center \citep{minor2017}. Since the Einstein radius of the lens typically corresponds to a few kiloparsecs in the lens plane, most of that distance lies 
along the line of sight to the lens plane, i.e. the plane containing the center 
of the host galaxy. Thus, in most cases the tidal radius will not affect the 
lensing dramatically, however we will investigate the effect of tidal 
truncation in detail in Section \ref{sec:tidal_truncation}.

Since the Hubble PSF is relatively undersampled compared to the pixel size, the 
images produced can be sensitive to how the ray-tracing is done, particularly 
near the critical curve. To mitigate this, we split each image pixel into 
N$\times$N subpixels and ray-trace the center point of each subpixel to the source 
plane, assigning it the surface brightness given by the source profile at the 
position of the ray-traced point. These surface brightness values are 
then averaged to find the surface brightness of the image pixel.  After the 
ray-tracing is complete, the resulting pixel values are then convolved with the 
PSF and Gaussian noise is added (similar to the noise level in SDSSJ0946+1006) 
to generate the mock image. An example with a $10^{10}M_\odot$ perturber is shown in Figure \ref{fig:mockdata}.  We find 
that 3$\times$3 splitting achieves sufficient accuracy without adding too much 
computational burden; a greater number of subpixels changes the surface 
brightness values very little while adding significant computational cost.

Our primary method for modeling the mock data uses a parameterized 
surface brightness profile, with the ray tracing as outlined above. This allows 
for relatively rapid exploration of the parameter space, which consists of both 
lens and source model parameters, hence reducing computational cost compared to 
reconstructing pixellated sources. However, for the purposes of constraining the detailed properties of subhalos, using an elliptical Gaussian profile would be too restrictive a prior, for two reasons: 1) actual galaxies rarely (if ever) have perfectly elliptical contours and Gaussian profiles; 2) even if the true source is Gaussian (as with our mock data), if one is modeling a subhalo with an incorrect mass or concentration, this systematic can often be (at least partially) absorbed into the inferred source galaxy by perturbing the isophotes or deviating from a Gaussian profile. This is also the case if the data are perturbed by a subhalo and one is not including a subhalo in the model at all. Thus, if one does not allow enough freedom in the source galaxy model, the constraints on concentration and mass may appear stronger than they really are.

For our modeling runs where we focus on constraining the concentrations of subhalos (Sections \ref{sec:mc_constraints} and \ref{sec:mc_constraints_hires}), we therefore adopt the following approach: we start with a Sersic profile where the isophotes are ``generalized ellipses'' with a radial coordinate defined by

\begin{equation}
    r_0(x,y) = \left(|x-x_0|^{C_0+2} + \left|\frac{y-y_0}{q}\right|^{C_0+2}\right)^\frac{1}{C_0+2}
\end{equation}

where $q$ is the axis ratio and $C_0$ is the ``boxiness'' parameter, such that $C_0=0$ corresponds to a perfectly elliptical profile. We then add Fourier mode perturbations to the isophotes, as follows:

\begin{equation}
    r(x,y) = r_0(x,y)\left\{1+\sum_{m=1}^N \left[a_m \cos(m\theta) + b_m\sin(m\theta)\right]\right\}
\end{equation}

From experimentation we find that parameter exploration can become difficult beyond 4-5 Fourier modes, and the $m=2$ mode is quite degenerate with the axis ratio parameter $q$. Thus, in our primary fits, we include the modes $m=1,3,4,5,6$, with sine and cosine terms for each. For high $m$ modes, fluctuations can easily become quite rapid, leading to noisy source solutions. To regulate this, we switch to the scaled amplitudes $\alpha_m = m a_m$, $\beta_m = m b_m$, which are the amplitudes of the azimuthal derivative $dr/d\theta$. By using these scaled amplitudes as free parameters, we can set an upper prior limit on the rate of change of the contours that applies equally to all modes. Our method for perturbing the isophotes is essentially identical to that employed by the GALFIT algorithm for fitting galaxy images \citep{peng2010}, except that instead of including a phase angle parameter, we use the scaled amplitudes for both sine and cosine terms as free parameters; this ensures that there is no coordinate singularity in the limit of very small amplitudes. While our source model may not exhibit all the freedom that a pixellated source does, it nonetheless allows for a wide range of source morphologies. 


The parameter exploration is done using the PolyChord algorithm, which is a variant of nested sampling that accommodates a large number of parameters and produces the Bayesian model evidence as well as posterior samples. In each case, the simulated image is first fit without a perturbing subhalo in the model, then the procedure is repeated with a subhalo included in the model. We then calculate the Bayes factor, defined as the ratio of  Bayesian evidences $K = \mathcal{E}_{sub} / \mathcal{E}_{no sub}$.  In practice 
we find that a subhalo is detected (such that the posterior includes the 
correct location of the subhalo) if $K \gtrsim 3$. However, in cases that are 
just above this threshold (up to $K \sim 10$ or so), there is a large 
uncertainty in the subhalo's position and mass, and in a few cases multiple 
modes exist, including one or more fictitious modes in addition to the correct 
one.  Moreover, in real life additional systematics may complicate detection by 
introducing false positives: the primary lens model may differ from the actual 
lens profile, e.g.  by having non-elliptical contours or twisted isodensity 
contours, and thus a fictitious subhalo might be preferred to make up these 
deficiencies. In addition, a more flexible source model (such as source pixel inversion; \citealt{suyu2006,vegetti2009}) may absorb small lensing perturbations into the source, resulting in a non-detection. With this in mind, for each mock data analysis, we conservatively 
call the results a detection if the Bayes factor $K$ is greater than 10.

\subsection{Mock data grid}\label{sec:mockdata_grid}

The density of subhalos is parameterized by the concentration $c_{200} 
\equiv r_{200}/r_s$ where $r_{200}$, $r_s$ are the NFW (approximate) virial radius and scale  radius respectively. \footnote{Although $r_{200}$ cannot be a reliable approximation to the virial radius of a tidally truncated subhalo, it is nevertheless a straightforward procedure to fit a subhalo using an NFW profile and then calculate $r_{200}$ and $m_{200}$, which is the virial radius and mass the subhalo \emph{would} have if it were actually a field halo at the given redshift. Thus a subhalo can be parameterized this way, as long as the parameters are interpreted correctly. This is the approach taken in recent lens modeling studies \citep{ritondale2019,despali2018} and in analyzing subhalo populations in N-body simulations \citep{moline2017}.}  When simulating different subhalo concentrations, we aim  to cover the expected scatter for a dark matter subhalo's given mass in $\Lambda$CDM simulations. For field halos, N-body simulations generally agree that dark matter halo concentrations follow a log-normal distribution with dispersion $\sigma_{\log c} \approx 0.11$ dex, although there has been slight variations in the median concentration of subhalos of given mass depending on the cosmological model adopted. These differences are generally smaller than the scatter, so for our purposes it is sufficient to adopt a single mass-concentration relation for our mock data, which we take from \cite{dutton2014}. Thus for  each subhalo, we will choose concentrations according to
\begin{equation}
\log c = \log\bar c(M,z) + \Delta_c\sigma_{\log c}
\label{logc_eq}
\end{equation}
where $\Delta_c$ will range from -2 up to 3. The high upper value for $\Delta_c$ is motivated by the fact that subhalos 
are expected to have slightly higher concentrations compared to field halos, 
since more dense subhalos are more likely to survive tidal stripping.

We justify this as follows. \cite{moline2017} found in the Via Lactea and Elvis simulations that the subhalo median concentration is approximately equal to the median value for field halos, multiplied by a factor $\mathcal{F} = (1+b\log(x_{sub}))$, where $b=-0.54$ and $x_{sub} = r_{sub}/r_{vir}$ is the ratio of the subhalo's distance from the host galaxy center to the host galaxy's virial radius. While it is unclear whether this factor differs significantly for large elliptical galaxy hosts, we can at least use it to guide our intuition about what concentrations may be expected among perturbing subhalos. If we consider subhalos of median concentration and use $\Delta_c\sigma_{\log c}$ as a proxy for the ``boost'' in concentration subhalos get over field halos, then we can set $\Delta_c\sigma_{\log c} \approx \mathcal{F}$ and find roughly what $x_{sub}$ corresponds to a given $\Delta_c$. For example, we find that for $\Delta_c = 1\sigma_{\log c}$ above the median field value, the subhalos are located at $\approx 0.3$ times the virial radius. Many subhalos are located at or within such a radius, so we can expect that such a boost in concentration is relatively common for subhalos; in addition, if one assumes that subhalos experience a similar scatter in concentration as field halos \citep{moline2017}, concentrations 2$\sigma$ above the median field value should not be uncommon among subhalos, with a small percentage reaching as high as 3$\sigma$ above the median. For $\Delta_c = 2\sigma_{\log c}$, one finds that subhalos are located at $\approx 0.06$ times the virial radius; only a very small fraction of subhalos lie within this radius. Again, given the scatter, it follows that in CDM only a very small fraction of subhalos will have concentrations $3\sigma$ higher than the median field value.

In view of the above considerations, our mock data grid consists of the following parameter values:

\begin{itemize}
	\item Lens redshift $z_{lens}$: 0.2, 0.5. We keep the primary lens's 
Einstein radius the same for either redshift (implying a somewhat more massive primary lens, by roughly a factor of 2.5, at the higher redshift). The source redshift is kept fixed at 
$z_{src}=2$.
	\item Subhalo (untruncated) mass $m_{200}$: $10^8M_\odot$, $10^9M_\odot$, $10^{10}M_\odot$.
	\item Subhalo concentration $c_{200}$, given by equation \ref{logc_eq} where 
$\Delta_c$ takes the following values: -2, -1, 0, 1, 2, 3.
	\item Subhalo's projected distance from the (unperturbed) critical curve: 
$-0.15''$, $-0.08''$, $-0.01''$, $0.06''$, $0.13''$ (where negative values are 
outside the critical curve, positive values inside). Note, these distances are 
measured with respect to the nearest point where the critical curve 
\emph{would} be if there were no perturbation present. 
	\item Source size, given by width of Gaussian surface brightness profile 
$\sigma_s$: $0.03''$, $0.06''$. We refer to these as ``small'' and ``large'' 
source galaxies respectively. To make a detection more likely, the small source 
is moved slightly closer to the caustic curve compared to the large source.
\end{itemize}

In total, we have a grid of 360 mock lenses. In all of our modeling runs, the 
position of the subhalo $x_{sub}$ and $y_{sub}$ and all of the primary lens parameters are varied freely. We will model the subhalo and source galaxy in three different ways:

\begin{itemize}
	\item \textbf{Method 1:} The mass $m_{200}$ and concentration $c_{200}$ are both varied during the fit, and the source is modeled with a Sersic profile with boxiness parameter and five Fourier modes (described in Section \ref{sec:lensmodel}). We focus on these modeling runs in Sections \ref{sec:mc_constraints} and \ref{sec:mc_constraints_hires}. Since the more flexible source model increases the computational burden, only a subset of the above mock data are modeled for these sections.
	
	\item \textbf{Method 2:} The mass $m_{200}$ and concentration $c_{200}$ are both varied during the fit, and the source is modeled with a Gaussian profile. We focus on these modeling runs to explore detectability of perturbations in Sections \ref{sec:detections}, and the degeneracy between mass and concentration in Section \ref{sec:mc_degeneracy}.
	
	\item \textbf{Method 3:} Only the mass $m_{200}$ is varied freely, while the concentration is always set to the median value $\bar c(M,z)$ during the fit. The source is modeled with a Gaussian profile. These fits are  explored in Sections \ref{sec:mbias} and \ref{sec:robust_mass}.
\end{itemize}

\begin{figure}[t]
	\centering
	\includegraphics[height=0.8\hsize,width=0.8\hsize]{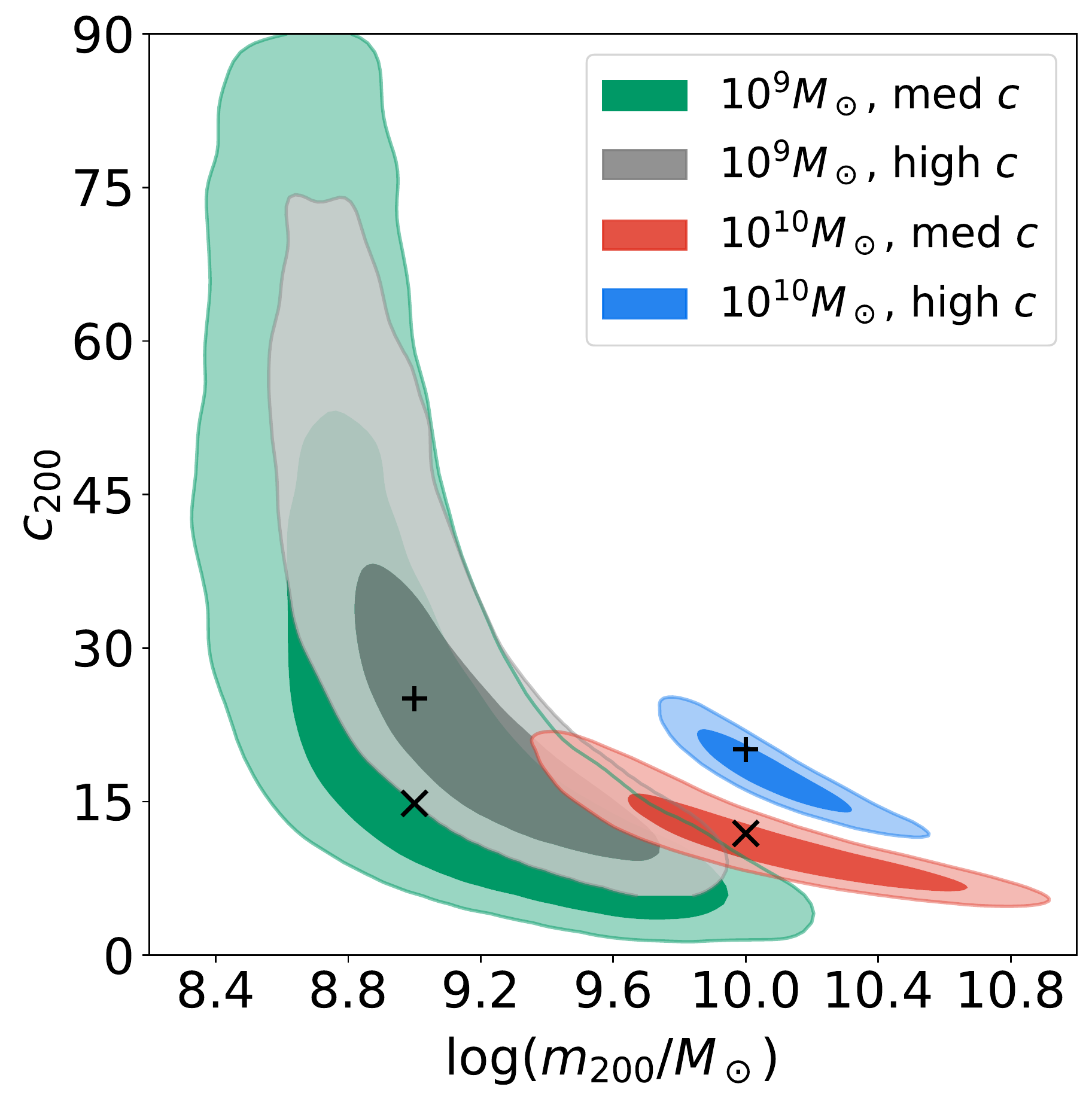}
\caption{Joint posteriors in log-mass $\log(m_{200}/M_\odot)$ and concentration $c_{200}$ of 
perturbing subhalos in four of our mock data fits at HST resolution and $z_{lens}=0.2$. Green and grey contours 
represent posteriors for $10^9M_\odot$ subhalos with concentrations at the 
median value $\bar c(M,z)$ and 2$\sigma$ above the median, respectively; red and blue contours represent $10^{10}M_\odot$ subhalos, again at the median and 
2$\sigma$ above the median respectively. The `x' markers denote the ``true'' 
values for these parameters for the median concentration cases, while the `+' 
markers denote the true values for the high concentration cases.}

\label{fig:mcplot}
\end{figure}

\begin{figure}[t]
	\centering
	\includegraphics[height=0.8\hsize,width=0.8\hsize]{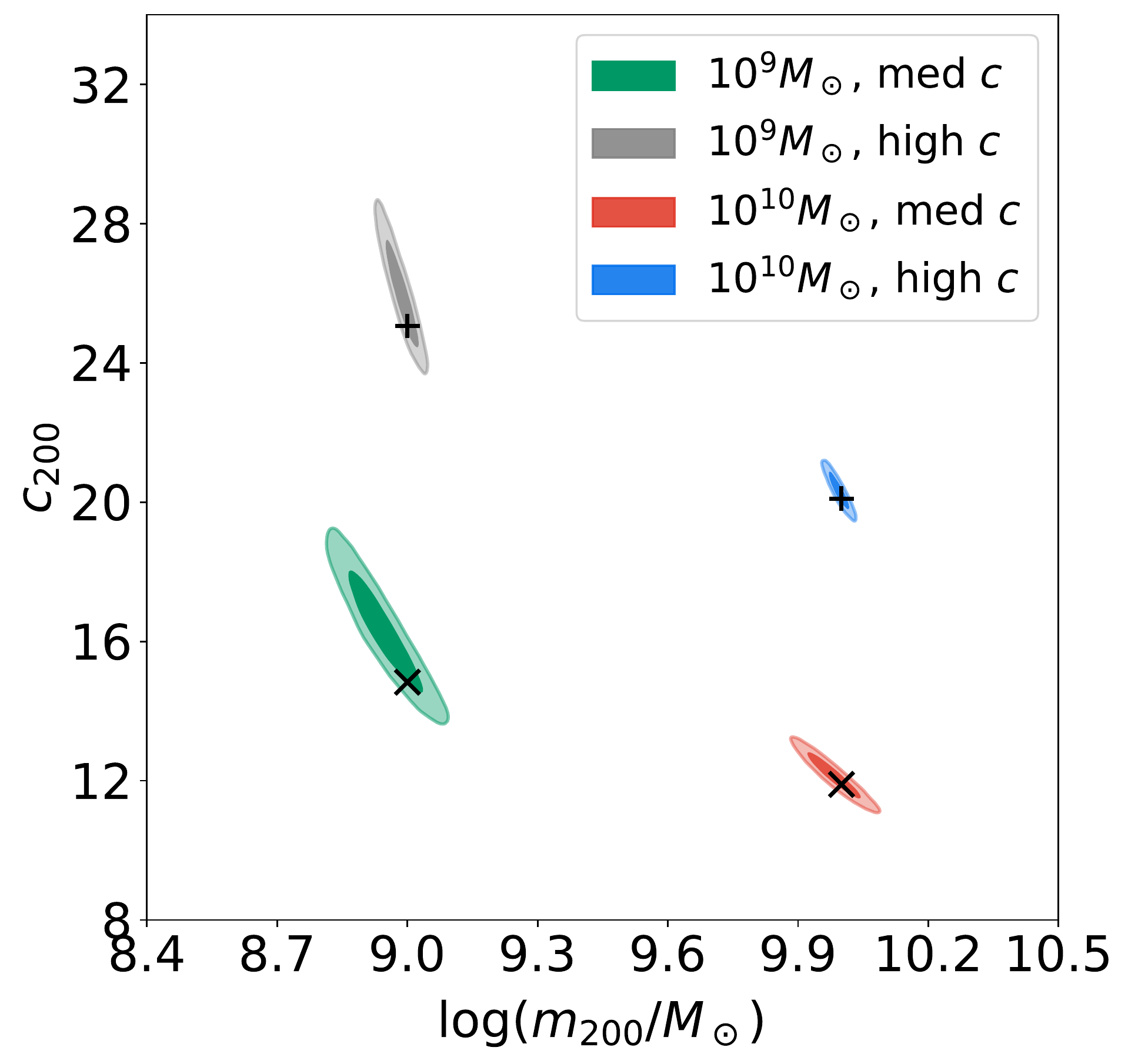}
\caption{Joint posteriors in log-mass $\log(m_{200}/M_\odot)$ and concentration $c_{200}$ of $10^8M_\odot$ and $10^9M_\odot$ perturbing subhalos, inferred using high-resolution simulated images with pixel size $\sim0.01''$ and PSF width equal to twice the pixel size. Green and grey contours represent posteriors for $10^8M_\odot$ subhalos with concentrations at 2$\sigma$ below the median value $\bar c(M,z)$ and equal to the median value, respectively; red and blue contours represent $10^{9}M_\odot$ subhalos, again at 2$\sigma$ below the median and at the median respectively.  The `x' markers denote the ``true'' values for these parameters for the median concentration cases, while the `+' markers denote the true values for the high concentration cases.}
\label{fig:mcplot_hires}
\end{figure}

In all modeling runs, the mass is allowed to vary from $10^6 M_\odot - 
10^{11}M_\odot$ and is parameterized as $\log m_{200}$ with a uniform prior 
over this range (this is equivalent to assuming a logarithmic prior in the 
subhalo's mass).  We use $\log(m_{200}/M_\odot)$ as our parameter to allow for better 
initial coverage of the parameter space during the nested sampling, since the  subhalo mass ranges over several orders of magnitude during the fit.  When 
varying the concentration, we use $c_{200}$ as our parameter directly with a 
logarithmic prior in this parameter covering the range (0.5,100). The host galaxy is modeled with an ellipsoidal power law profile, where all parameters are varied: $b$, $\alpha$, $q$, $\theta$, $x_c$ and $y_c$ (the Einstein radius, density log-slope, axis ratio, orientation angle, and center coordinates respectively).

\section{Constraints on subhalo concentration and mass}\label{sec:results}

\subsection{Mass-concentration constraints at HST resolution}\label{sec:mc_constraints}

Since most $10^9M_\odot$ subhalos and all $10^{10}M_\odot$ subhalos are 
detected with high significance in our modeling runs, we now investigate the 
resulting constraints in both mass and concentration for some representative 
cases. In Figure \ref{fig:mcplot} we plot joint posteriors in $\log(m_{200}/M_\odot)$ 
and $c_{200}$ for subhalos at $z_{lens}=0.2$ placed at the closest point to the 
critical curve (just $0.01''$ outside the unperturbed critical curve), for the ``large'' source galaxy cases.  Posteriors are plotted for $10^9M_\odot$ and $10^{10}M_\odot$ subhalos with concentrations at the median value $\bar c(M,z)$ and 2$\sigma$ above the median in each case.

In all cases, a degeneracy exists between mass and 
concentration, since a perturbation of similar size can be produced by 
increasing the mass while decreasing the concentration, or vice versa (this relationship will be explored in detail in Section \ref{sec:mc_degeneracy}).  
Although the constraints on concentration are weak for $10^9M_\odot$ subhalos, there is nevertheless a lower bound on concentration, and this was generally true for all of the fits that satisfy our detection criterion (Section \ref{sec:mockdata}).  Nevertheless, impressive constraints on the concentration can be obtained for $10^{10}M_\odot$ subhalos.  
In the median concentration case for which $c_{true}=11.9$, we infer a 
50th percentile value $c_{fit} = 14.9_{-5.8}^{+8.1}$ where the uncertainties 
give the 95\% credible interval.  For the high concentration case where 
$c_{true}=20.1$, we infer $c_{fit} = 22.3_{-4.5}^{+5.6}$. Remarkably, these 
uncertainties are smaller than the expected 2$\sigma$ scatter in concentration for $10^{10}M_\odot$ halos. In principle, it follows that even with a small number of such detections at HST resolution, one may be able to distinguish between the mass-concentration relation expected in CDM (for subhalos) versus alternate scenarios such as warm or self-interacting dark matter (if one assumes the standard power-law spectrum of inflationary perturbations).

All of the cases shown in Figure \ref{fig:mcplot} are for subhalos whose 
projected position is just outside the critical curve (by $0.01''$). To 
investigate the constraints for subhalos further away from the critical curve, 
in Figure \ref{fig:mcplot_m2} we plot the resulting constraints for a 
high-concentration $10^{10}M_\odot$ subhalo at different positions: the grey 
contours correspond to a subhalo at the innermost position, $0.13''$ inside the 
critical curve; red contours are for a subhalo close to the critical curve 
(same as blue contours in Figure \ref{fig:mcplot}); blue contours are for a 
subhalo at the outermost position, $0.15''$ outside the critical curve.  
Interestingly, the constraints for the outermost subhalo (blue) are nearly as 
good compared to the subhalo close to the critical curve, whereas the 
uncertainties are significantly weaker for the innermost subhalo. Nevertheless, 
significant constraints are obtained in each of these cases.


\subsection{Mass-concentration constraints at high 
resolution}\label{sec:mc_constraints_hires}

Although the constraints on concentration are weak for $\lesssim 10^9M_\odot$ 
subhalos at HST resolution, we can ask whether better constraints will be 
possible for higher resolutions, e.g. using long baseline interferometry or next-generation very 
large telescopes. To investigate this, we simulate lenses similar to those 
shown in Figure \ref{fig:mcplot}, but with $10^8M_\odot$ and $10^9M_\odot$ 
subhalos, imaged with pixels four times smaller than HST ($\sim0.013''$ compared to $\sim0.049''$ for HST) and 
assuming a Gaussian PSF whose width is twice the pixel size. Note that both the 
Thirty Meter Telescope and Giant Magellan Telescope will exceed this 
resolution; in principle the ALMA radio telescope array can reach this resolution 
at long baseline, although the pixel noise is highly correlated due to the 
interferometry.

The high-resolution constraints on concentration and mass are shown in Figure 
\ref{fig:mcplot_hires}. Remarkably, at this resolution the concentration for 
$10^9M_\odot$ subhalos can be well constrained even for low concentrations. At 
median concentration with $c_{true}=14.83$, we infer 
$c_{fit}=15.78_{-2.08}^{+2.31}$, much smaller than the expected scatter in 
concentration at this mass for $\Lambda$CDM. Even more remarkably, meaningful 
constraints are obtained for $10^8M_\odot$ even at median or low concentrations 
whereas neither of these subhalos were even detected to high significance at 
the HST resolution (in fact the low-concentration perturbation was not detected 
at all). The latter results should be interpreted with some caution, since it is possible that if the source model is allowed more freedom than in our model (e.g. for pixellated sources), perturbations as small as $10^8M_\odot$ might be at least partially absorbed into the reconstructed source, possibly weakening the constraints; as pixellated source inversions are computationally intensive, we do not investigate this possibility here. Nevertheless, we conclude that significant constraints on the mass-concentration of 
perturbing subhalos, possibly down to $10^8M_\odot$, will be possible for next generation 
telescopes that attain resolutions $\sim 0.01''$ or better.

\begin{figure}
	\centering
	\includegraphics[height=0.7\hsize,width=0.9\hsize]{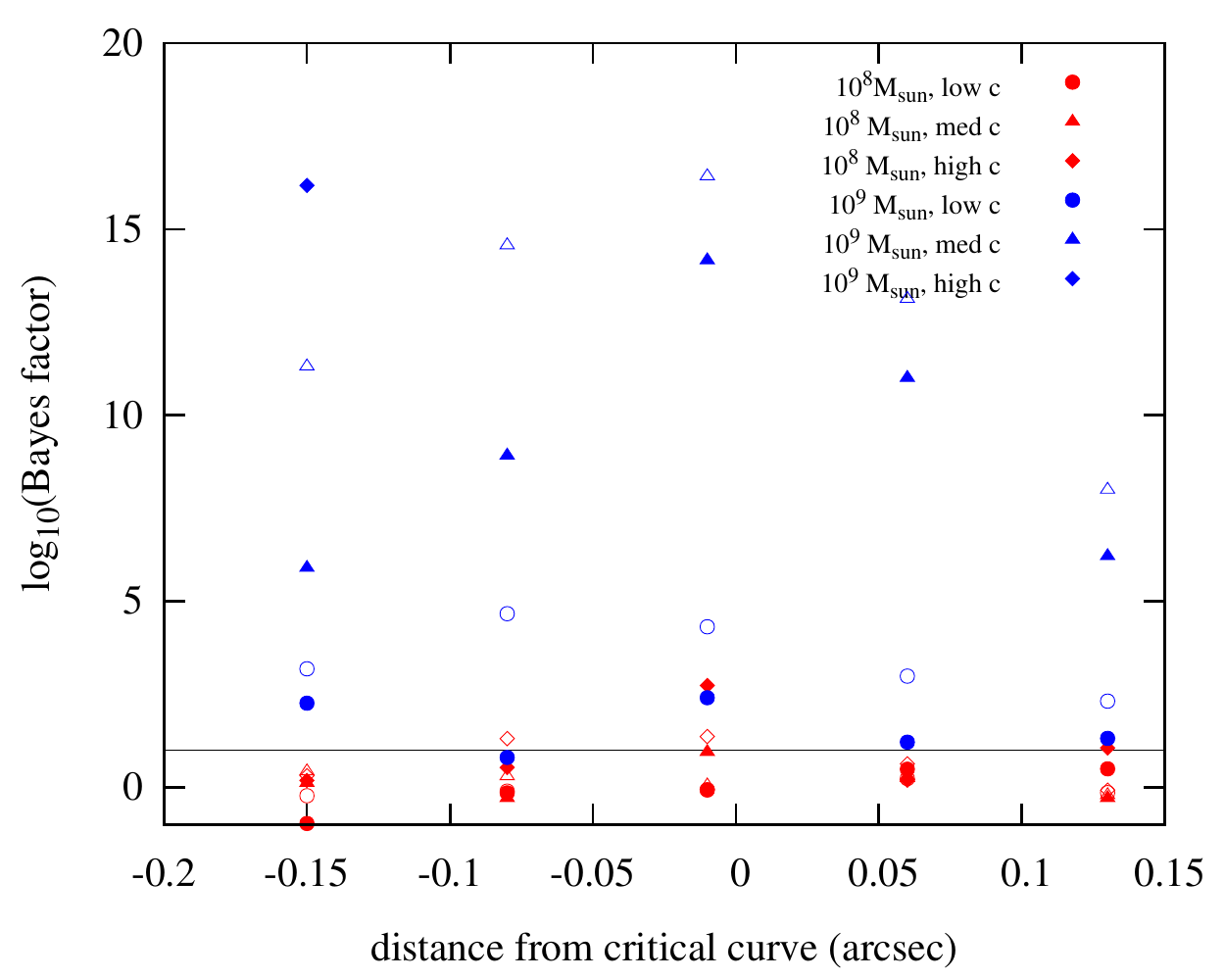}
\caption{Log-Bayes factors $\log_{10}K$ versus distance from the critical curve 
for the subset of mock data at $z_{lens}=0.2$, shown up to $\log_{10}K = 20$ 
(all $10^{10}M_\odot$ subhalos exceed this range and therefore are not shown).  
The black line represents the threshold for which we consider a subhalo 
``detection'' to have occurred, corresponding to $\log_{10}K \gtrsim 1$.  Filled 
markers denote the lenses for which the ``small'' source galaxy is being 
lensed, while open markers correspond to the ``large'' source being lensed 
(with unlensed widths given by $0.03''$ and $0.06''$ respectively). The 
distances are measured from the nearest point on the \emph{unperturbed} 
critical curve, with negative values lying outside the critical curve and vice 
versa.}
\label{fig:detections}
\end{figure}

\section{Detections}\label{sec:detections}

To examine which subhalo perturbations are actually detectable, in Figure 
\ref{fig:detections} we plot the logarithm of the Bayes factor, $\log_{10}K$, 
against the subhalo distance from the (unperturbed) critical curve for a subset 
of lenses from our mock data runs. The dashed line represents our detection 
criterion, namely $\log K > 1$; all cases where $\log K$ is greater than 20 are not 
shown in the plot (this includes all subhalos with mass $10^{10}M_\odot$).  Red 
markers denote $10^8M_\odot$ subhalos, while blue markers denote $10^9M_\odot$ 
subhalos.  We only show the subset of concentrations with $\Delta_c = 
(-2,0,2)$, (i.e.  2$\sigma$ below the median, at the median, or 2$\sigma$ above 
the median); these are denoted by circles, triangles, and diamonds 
respectively.  Filled markers denote lenses generated using the ``small'' 
source galaxy, while open markers correspond to the ``large'' source galaxy.

Note that nearly all $10^9M_\odot$ subhalos are unambiguously detected, with 
the exception of the low-concentration subhalos with small sources---these 
hover near the threshold of detectability, unless very close to the critical 
curve.  This does not mean that the subhalo does not perturb these small 
sources noticeably (it certainly does), but rather the perturbation is small 
enough that it can be degenerate with the parameters governing the primary 
lens. In addition, because the primary lens parameters are less well 
constrained with the small source, there is more freedom to vary them and 
possibly mimic the subhalo perturbation. However, we can conclude that all 
$10^9M_\odot$ perturbations with large sources and/or concentrations at or 
above the median are apparently detectable at HST resolutions, at least if they lie within 
$\sim 0.1$'' from the critical curve. One important caveat should be given here, however: for the fits in this section, the source was modeled as a Gaussian without Fourier modes. If more freedom is allowed to the source (e.g. with Fourier modes, or with a pixellated source), it is possible that a few of these cases may be well-modeled without a subhalo, since the perturbation may be absorbed into the source itself while still achieving a good fit.

For $10^8M_\odot$ the story is very different: at most a few subhalos were 
unambigiuously detected, with the clearest detection occurring when the subhalo 
has a high concentration, is perturbing a small source, and is very close to 
the critical curve. In general, the smaller source galaxy is more favorable for 
detection due to the small size of the perturbation: a rapid variation in 
surface brightness is required for the perturbation to be noticeable at all, 
unless the concentration is quite high (as the open diamonds show). We conclude that relatively few (if any) $10^8M_\odot$ subhalos will be detected at HST 
resolutions in CDM, and those that are detected are very likely to have high 
concentrations; again however, it is possible that with a more flexible source model, even these cases may be well-modeled without a subhalo.

At the higher lens redshift, $z_{lens}=0.5$, the results are essentially the same, hence we omit the figure for this case. We note however that most of the Bayesian evidences are slightly lower compared to 
$z_{lens}=0.2$. Indeed, nearly all of the subhalo perturbations are less 
pronounced at higher redshift, due to the fact that the angular size of the NFW 
scale radius is smaller at higher redshift. Since the ratio of angular diameter 
distances $D_{lens}(z=0.5)/D_{lens}(z=0.2) \approx 0.5$, the scale radius at 
the higher redshift appears smaller by nearly this factor (although not 
precisely, since halos have lower concentrations at higher redshift).  Thus the 
noticeable extent of a subhalo's perturbation is smaller at high redshift, at 
least for a primary lens of fixed Einstein radius.  As a result, \emph{none} of 
the $10^8M_\odot$ subhalos were unambiguously detected at $z_{lens}=0.5$, even 
if placed very close to the critical curve.  Nevertheless, nearly all 
$10^9M_\odot$ subhalos are detected, with the exception of the 
low-concentration + small source cases with subhalo placed at the extreme 
positions inside or outside the critical curve.

These results imply that with HST-like resolution, perturbing subhalos of mass 
$< 10^9M_\odot$ are more likely to be detected if they have a concentration at 
or above the median expected value for $\Lambda$CDM. For $\sim 10^8M_\odot$ 
subhalos, \emph{only} high-concentration subhalos have a chance to be detected, 
and only if they are quite close to the critical curve; for $z_{lens} \gtrsim 
0.5$, detection of such low-mass subhalos is unlikely in any configuration.

\begin{figure}
	\centering
	\includegraphics[height=0.8\hsize,width=0.8\hsize]{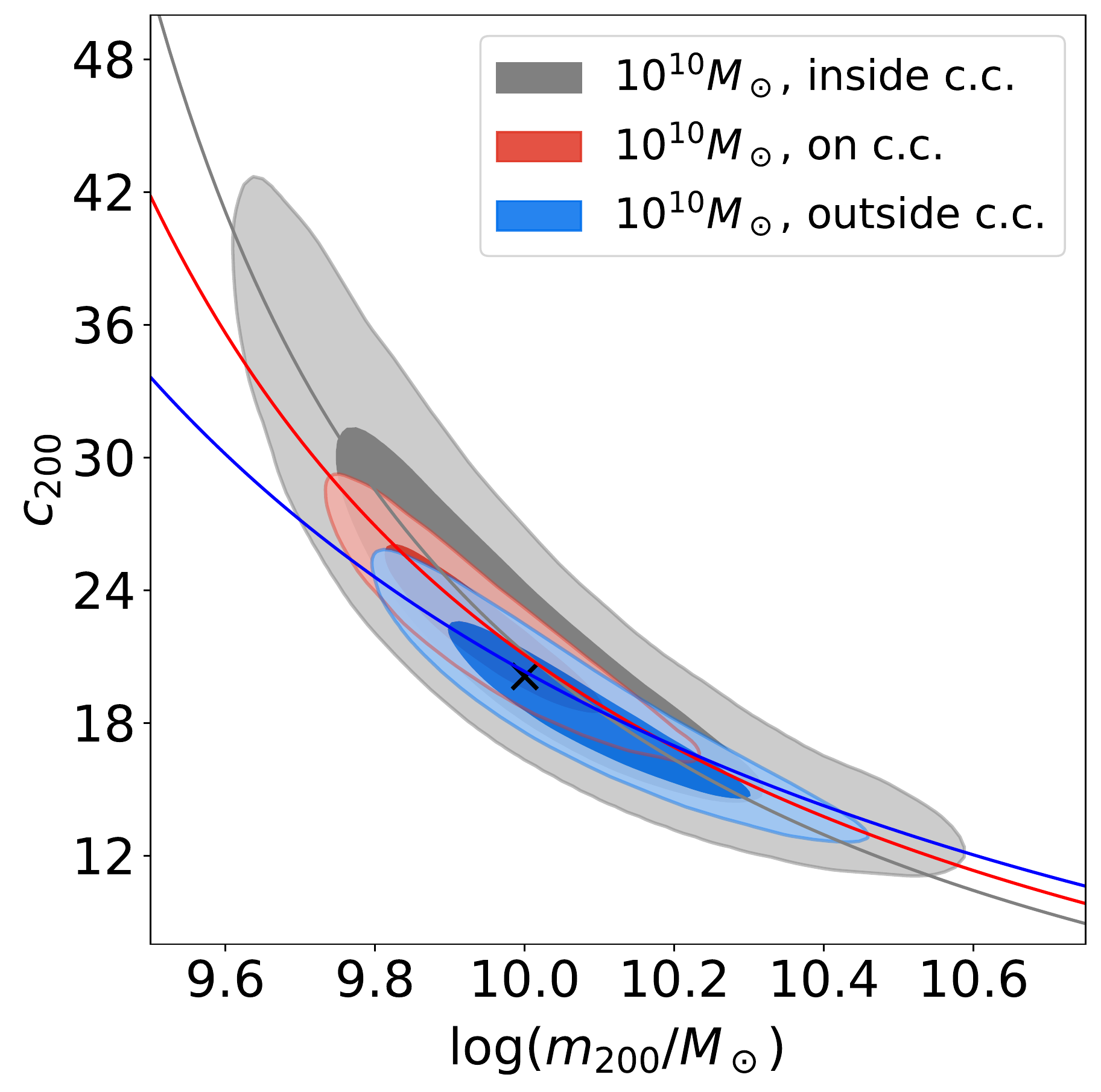}
\caption{Joint posteriors for a $10^{10}M_\odot$ perturbing subhalo, at different projected positions with respect to the critical curve (and $z_{lens}=0.2$). Grey, red and blue contours 
correspond to subhalos located $0.13''$ inside, $0.01''$ outside, and $0.15''$ 
outside the critical curve respectively. The `x' marks the true values for 
these parameters. The solid curves represent the set of points 
($\log(m_{200})$,$c_{200}$) for which the subhalo contains the same projected mass 
within the perturbation radius $r_{\delta c}$ of the best-fit model in each 
case.}
\label{fig:mcplot_m2}
\end{figure}

\section{Physical interpretation of the degeneracy between concentration and 
mass}\label{sec:mc_degeneracy}

Note that the correlation in the inferred concentration versus mass is slightly 
different in the posteriors for each subhalo position in Figure 
\ref{fig:mcplot_m2}, with noticeably different ``tilt''. What determines this 
correlation? In \cite{minor2017} we showed that for subhalo perturbations, a 
characteristic perturbation radius can be defined, within which the subhalo's 
projected mass is determined robustly, provided the log-slope of the primary galaxy's density profile
is also well-determined. This subhalo perturbation radius $r_{\delta c}$ is 
defined as the projected distance from the subhalo's center to the point where 
the critical curve is perturbed the most, and to good approximation typically 
lies along the line from the primary galaxy's center to the position of the 
subhalo.  To check whether this mass is robustly determined for the 
$10^{10}M_\odot$ case plotted here, we used the following procedure: first, we 
calculated $r_{\delta c}$ for the best-fit model in each case, and the 
corresponding mass within this radius; next, for an array of $m_{200}$ values 
over the range of the posteriors, we calculated what concentration is required 
to keep the mass within $r_{\delta c}$ constant, using a numerical root finder.  
The resulting curves are plotted in Figure \ref{fig:mcplot_m2}, with colors matching the 
corresponding posterior.  Note that the overall correlation is recovered very 
well, indicating the perturbation radius is well constrained; the tilts differ 
because the inferred $r_{\delta c}$ is different in each case, due to the 
differing subhalo positions.

\begin{figure}
	\centering
	\includegraphics[height=0.9\hsize,width=0.9\hsize]{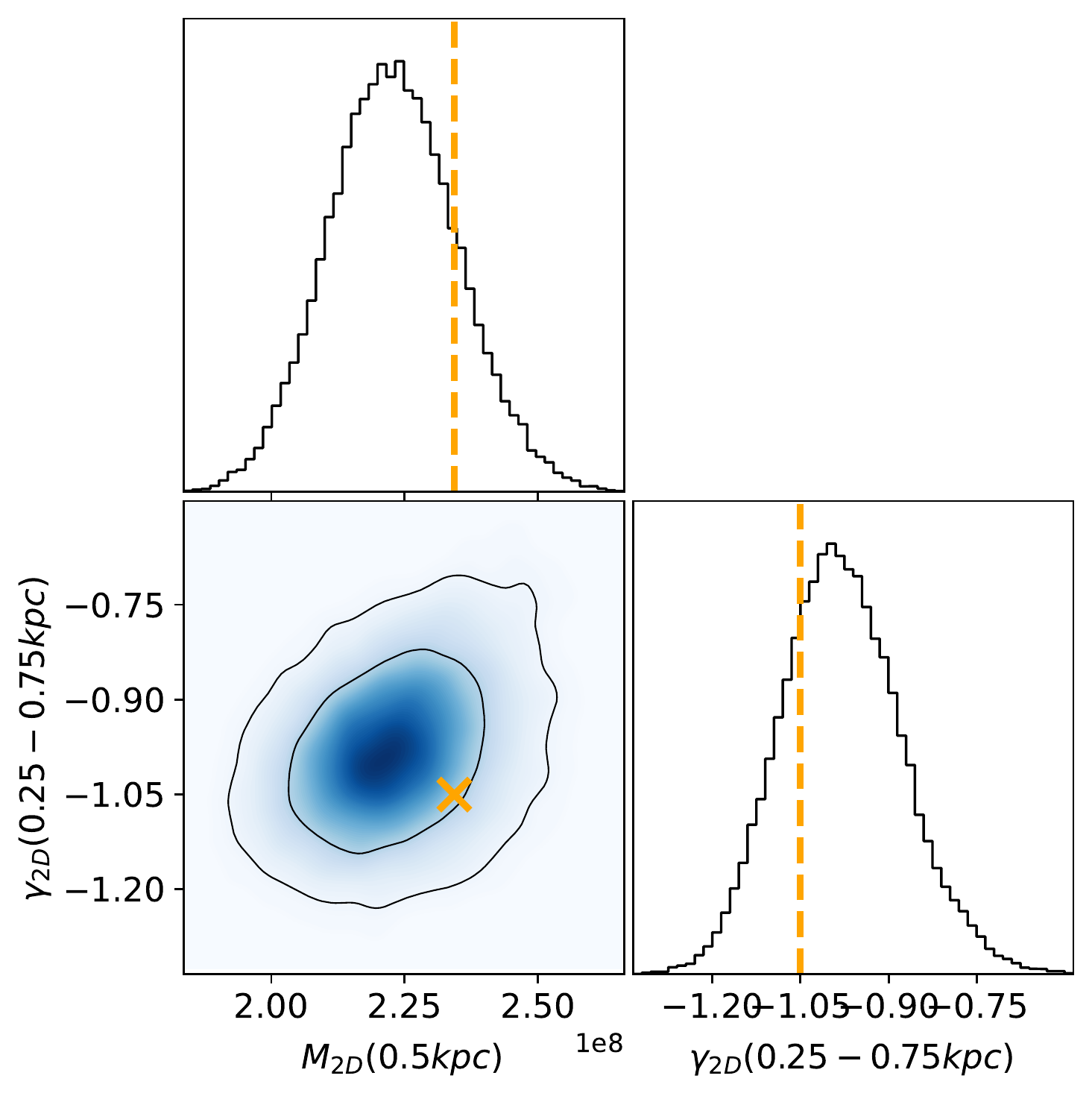}
\caption{Joint posterior in the projected subhalo mass within 0.5 kpc, $M_{2D}($0.5kpc$)$, versus the average log-slope of the density profile from 0.25 to 0.75 kpc, $\gamma_{2D}($0.25-0.75kpc$)$. The mock data was produced with a perturbing subhalo of mass $m_{200} = 10^{10}M_\odot$ and concentration 2$\sigma$ above the median CDM value, whose posterior in $m_{200}$ and $c_{200}$ is shown as the blue contour in Figure \ref{fig:mcplot}. After fitting the mock data, the subhalo's inferred perturbation radius $r_{\delta c} \approx$ 0.5 kpc, so we choose the projected mass and log-slope around this radius as our derived parameters. The actual parameter values are given by the orange marker. Note that, unlike $m_{200}$ and $c_{200}$, there is no significant degeneracy between these two parameters and the mass is well-constrained to within $\approx 10\%$.}
\label{fig:mpert_vs_logslope}
\end{figure}

Since the subhalo's projected mass within the perturbation radius can be determined robustly, another way to cast the results is to infer a derived parameter $M_{2D}(r_{\delta c,bf})$ where $r_{\delta c,bf}$ is the approximate best-fit perturbation radius from the fit. Likewise, as a proxy for concentration, one can define a derived parameter $\gamma_{2D}$ as the average density log-slope in the vicinity of the perturbation radius, since this is the region where the subhalo's perturbation should yield the most information about the profile. For example, in Figure \ref{fig:mpert_vs_logslope} we plot joint posteriors for the perturbing subhalo of mass $m_{200} = 10^{10}M_\odot$ and concentration $2\sigma$ above the median CDM value (whose posterior in $m_{200}$ and $c_{200}$ is shown as the blue contour in Figure \ref{fig:mcplot}). For this subhalo, the best-fit perturbation radius $r_{\delta c} \approx 0.5''$, so we choose this radius to evaluate $M_{2D}($0.5kpc$)$. We likewise define the average log-slope of the density profile from 0.25 to 0.75, $\gamma_{2D}(0.25-0.75$kpc$)$. Note that there is almost no discernible degeneracy between these two parameters, and $M_{2D}$ is well-constrained to within $\approx 10\%$. These inferences can be compared to dark matter simulations provided the simulation resolution is sufficiently high. Note that, unlike $M_{2D}$, there is no guarantee that the inferred log-slope will be independent of the subhalo profile chosen; its robustness may depend on the actual interval chosen for evaluating the slope. Nevertheless, the advantage of switching to log-slope is that it is more straightforward to compare to simulations and among different lensing solutions, as it does not depend on defining a theoretical $r_{200}$ or $m_{200}$ for a subhalo. In a companion paper (Minor et al. 2020, in prep), we use exactly this approach when modeling the subhalo detected in the gravitational lens SDSSJ0946+1006 \citep{vegetti2010}.

Although we have explained how the degeneracy between subhalo mass and concentration arises, we must still understand what determines the 
inferred limits on the concentration.  In Section \ref{sec:c_limits} we 
will investigate what is physically constraining the subhalo concentration at the high and low ends, and whether tidal truncation significantly alters these constraints.

\begin{figure*}
	\centering
	\subfigure[$10^{10}M_\odot$ subhalo]
	{
		\includegraphics[height=0.36\hsize,width=0.31\hsize]{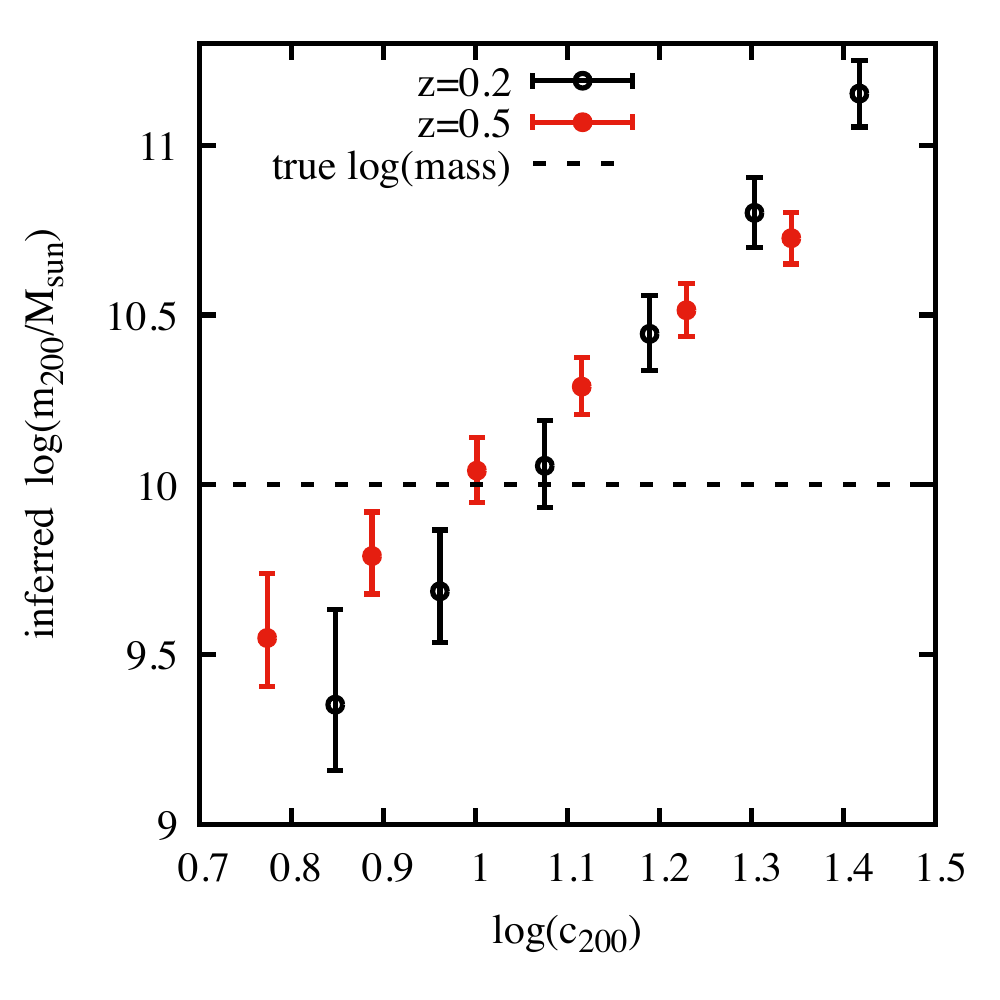}
		\label{fig:mbias_1e10}
	}
	\subfigure[$10^{9}M_\odot$ subhalo]
	{
		\includegraphics[height=0.36\hsize,width=0.31\hsize]{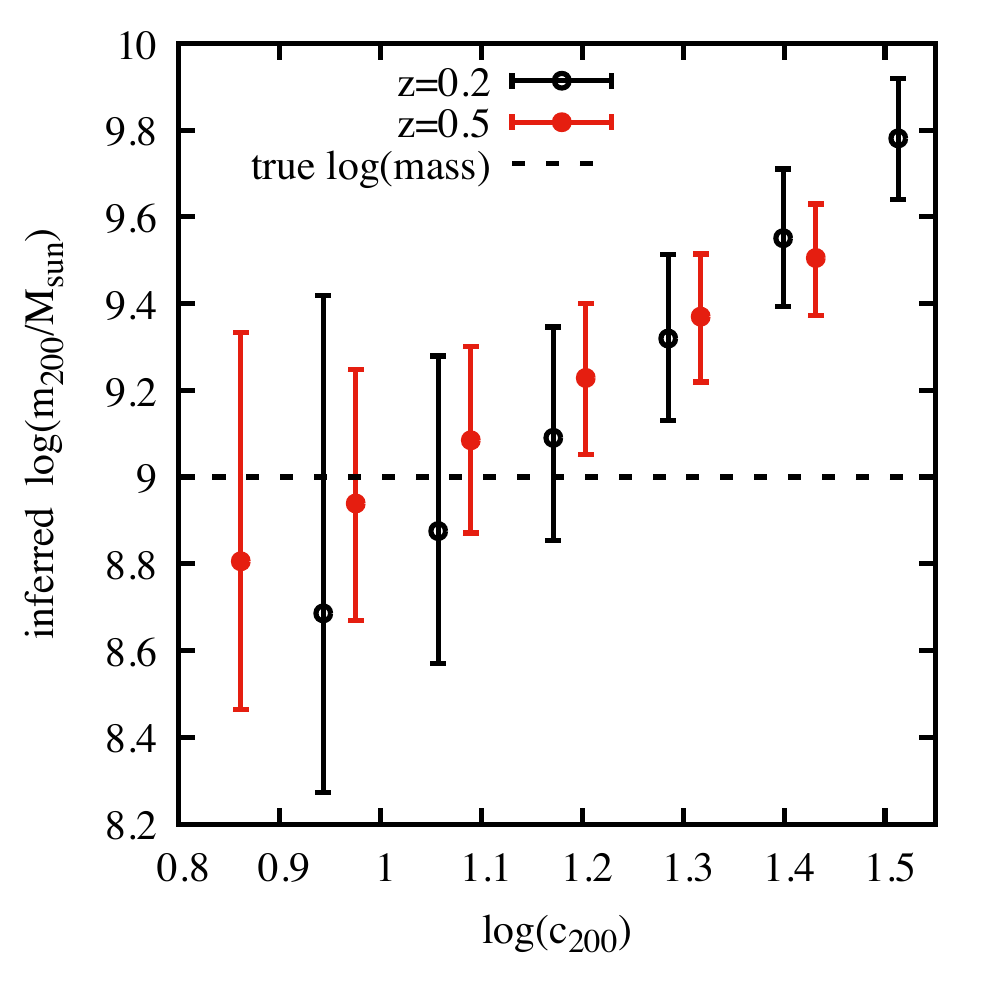}
		\label{fig:mbias_1e9}
	}
	\subfigure[$10^{8}M_\odot$ subhalo]
	{
		\includegraphics[height=0.36\hsize,width=0.31\hsize]{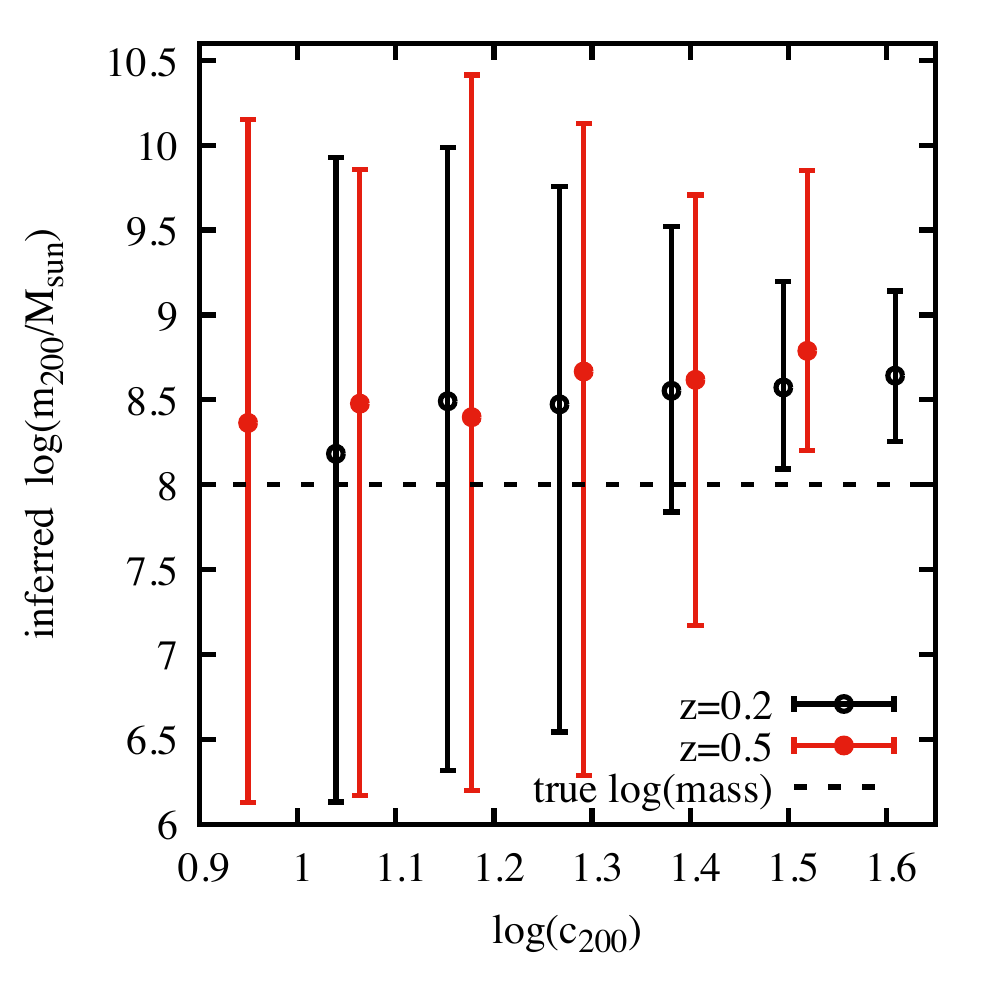}
		\label{fig:mbias_1e8}
	}
	\caption{Inferred subhalo mass $m_{200}$ plotted against the subhalo's 
actual (log) concentration $c_{200}$, where the modeling assumes the subhalo's 
concentration to be at the median value for $\Lambda$CDM during the fit. Black 
(red) points show the 50th percentile inferred value for $m_{200}$ for lenses 
at $z_{lens}=0.2$ ($z_{lens}=0.5$), while error bars show the 95\% credible 
intervals. Dashed line shows the subhalo's true mass in each case. The 
log-concentrations chosen for the mock data (horizontal axis) differ from the 
median value $\log\bar c(M,z)$ by (-2$\sigma$, -1$\sigma$, 0, 1$\sigma$, 
2$\sigma$, 3$\sigma$) where $\sigma \approx 0.11$ is the scatter in 
concentration observed in $\Lambda$CDM simulations.}
\label{fig:mbias}
\end{figure*}

\begin{figure*}
	\centering
	\subfigure[robust mass, $10^{10}M_\odot$]
	{
		\includegraphics[height=0.36\hsize,width=0.31\hsize]{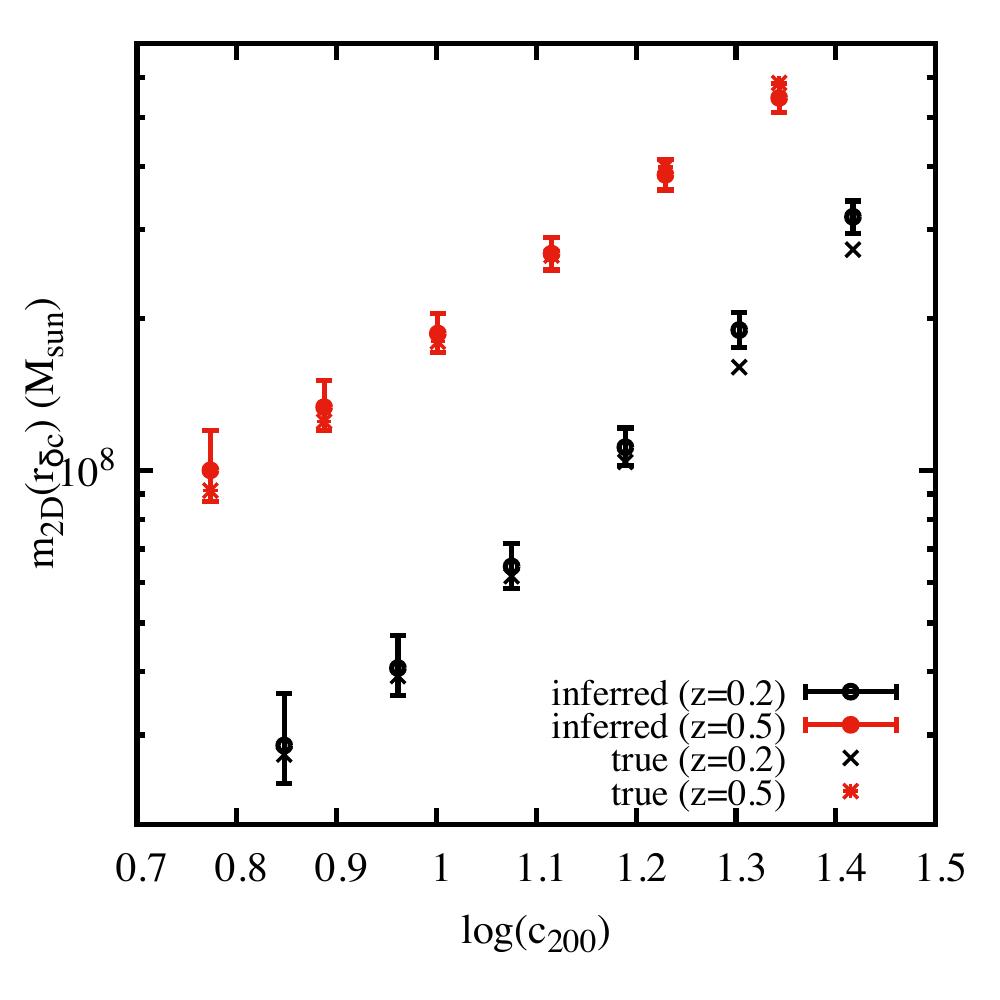}
		\label{fig:mpert_1e10}
	}
	\subfigure[robust mass, $10^{9}M_\odot$]
	{
		\includegraphics[height=0.36\hsize,width=0.31\hsize]{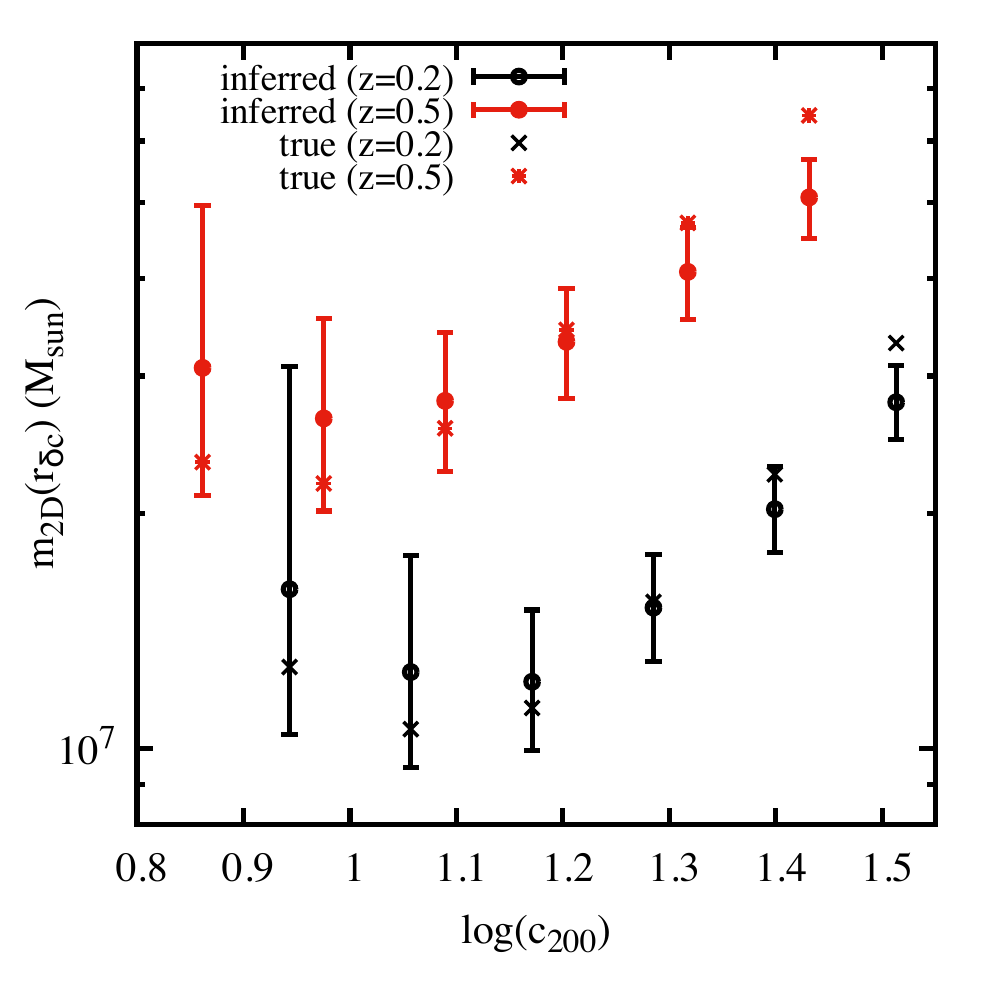}
		\label{fig:mpert_1e9}
	}
	\subfigure[robust mass, $10^{8}M_\odot$]
	{
		\includegraphics[height=0.36\hsize,width=0.31\hsize]{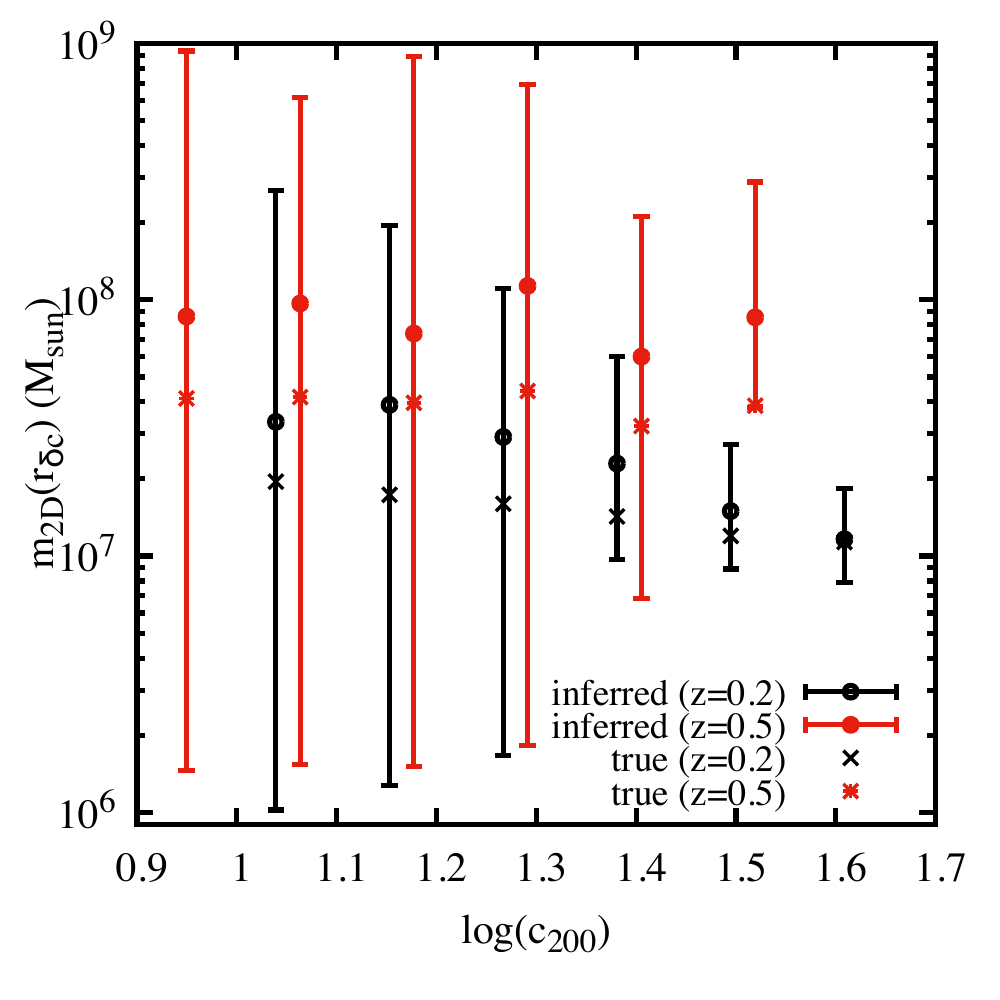}
		\label{fig:mpert_1e8}
	}
	\caption{Constraint on the subhalo's projected mass enclosed within its 
perturbation radius $r_{\delta c}$, plotted against the subhalo's actual (log) 
concentration $c_{200}$ where the modeling assumes the concentration to be at the 
median value for $\Lambda$CDM during the fit. Plotting conventions are the same 
as in Figure \ref{fig:mbias}, except that the corresponding true values 
$M_{2D,true}$ are plotted with a black `x' for $z=0.2$, and a red asterisk for 
$z=0.5$.
}
\label{fig:mpert}
\end{figure*}

\section{Can scatter in the mass-concentration relation be ignored when 
modeling subhalo perturbations?}\label{sec:mc_scatter}

\subsection{Bias in the inferred subhalo mass}\label{sec:mbias}

Several recent lens modeling studies have modeled subhalo perturbations using 
an NFW profile, under the assumption that the concentration takes on the median 
value for $\Lambda$CDM, ignoring the expected scatter in subhalo 
concentrations \citep{despali2018,ritondale2019,li2017}. However, Figure \ref{fig:mcplot} already suggests that the effect can 
be substantial: for example, if a $10^{10}M_\odot$ subhalo has a concentration 
2$\sigma$ above the median (the rightmost `x' in the figure), but one assumes 
that it has a median concentration (which is a factor of $\sim$1.7 smaller), 
the inferred mass indicated by the posterior will be substantially larger. The reason is straightforward: 
if one underestimates the concentration, then 
more mass is required to generate a sufficiently large perturbation (this is discussed in detail in Section \ref{sec:mc_degeneracy}).

Here we investigate to what extent the inferred mass of a subhalo can be biased 
due to scatter in concentration, if one assumes the subhalo's concentration is 
determined by the median value $\bar c(M,z)$ expected in $\Lambda$CDM. We model 
all of the mock data described in Section \ref{sec:mockdata_grid}, varying the 
subhalo's position and $\log(m_{200})$ parameters while keeping the concentration set to 
$\bar c(m_{200},z_{lens})$ during the fit.

In Figure \ref{fig:mbias} we plot the inferred subhalo mass $m_{200}$ versus 
the subhalo's actual concentration (or rather, log-concentration) for each of 
the subhalo masses in our mock data grid ($10^{10}M_\odot$, $10^9M_\odot$, 
$10^8M_\odot$ correspond to \ref{fig:mbias_1e10}, \ref{fig:mbias_1e9}, 
\ref{fig:mbias_1e8} respectively). The circles show the 50th percentile inferred 
values, while the error bars show the 95\% credible interval; black bars are 
for $z_{lens}=0.2$, while red bars are for $z_{lens}=0.5$. In each subplot, 
recall that the actual concentrations range from being -2$\sigma$ below the 
median value, all the way up to 3$\sigma$ above the median value, with the 
third bar from the left corresponding to subhalos at the median concentration.  
The subhalo was located near the critical curve (at $0.01''$ outside) for all 
the cases plotted. For $10^{10}M_\odot$ and $10^9M_\odot$, the ``large'' source 
galaxy was used since the uncertainties are somewhat smaller in these cases, 
whereas for $10^8M_\odot$, the small source was used since a greater number of 
detections occurred in this case.

In Figure \ref{fig:mbias_1e10}, one can see that for lenses at $z=0.2$, 
$10^{10}M_\odot$ subhalos that are 1$\sigma$ above the median have an inferred 
mass that is biased high by nearly a factor of 3, while subhalos 1$\sigma$ 
below the median are biased low by a factor of $\sim$2. At 2$\sigma$ the bias 
is even more severe, roughly a factor of $\sim$6. The corresponding bias at $z=0.5$ is somewhat smaller, roughly half as large (a factor of $\sim$3) for concentrations $2\sigma$ above the median; this pattern is similar for the $10^9M_\odot$ case. For $10^{9}M_\odot$ 
perturbers (Figure \ref{fig:mbias_1e9}) the bias is not as severe, but 
nevertheless $m_{200}$ is biased by a factor of $\sim$2 for subhalos whose 
concentrations are 1$\sigma$ above the median, and between 3-4 for subhalos 
2$\sigma$ above the median. For $10^8M_\odot$, only the subhalos above the 
median concentration were detected at all; in these cases however, we see bias 
of roughly a factor of 3, albeit with large uncertainties.

Although one might expect that very few subhalos ($\sim$5\%) would happen to 
have concentrations 2$\sigma$ away from the median, and hence the mass is 
unlikely to be biased by a factor greater than 2, this may not be the case, for 
two reasons. 1) Subhalos tend to have systematically higher concentrations than 
field halos, as a result of tidal interactions with the host galaxy; hence, a 
greater proportion of subhalos at high concentration is likely. 2) For subhalos 
of mass $\lesssim 10^9M_\odot$, high-significance detections are more likely to 
occur for subhalos whose concentrations are above the median value. Thus, even 
in $\Lambda$CDM one may expect that the inferred subhalo mass may be biased 
significantly in many cases, possibly up to a factor of 6, if scatter in concentration is neglected when modeling lenses.

Finally, an additional source of bias in the inferred perturber mass comes from the fact that many perturbing halos may actually be unassociated halos along the line-of-sight to the lens, rather than subhalos  \citep{xu2012,li2017,sengul2020}. If the redshift of the perturber differs significantly from the redshift of the primary lens, this may bias the inferred halo mass if unaccounted for during the lens modeling \citep{despali2018}. In principle this bias can be quantified using the same methodology in this paper, with a corresponding redshift scaling of the inferred mass (especially the projected mass $M_{2D}(r_{\delta c})$) depending on the unknown perturber redshift. We leave this analysis to future work.

\subsection{Testing the robust subhalo mass estimator}\label{sec:robust_mass}

If subhalos are indeed well-approximated by a spherical NFW profile with only 
modest tidal truncation in most cases, then a straightforward way to overcome 
the bias in the inferred mass is to vary \emph{both} the mass and concentration 
as free parameters. In fact, we have seen in Section \ref{sec:mc_constraints} 
that this approach can yield important constraints about the concentration 
itself.  Nevertheless, it is useful to have a mass estimate that is robust even 
if the assumed density profile is incorrect (e.g. due to incorrect assumptions 
about the concentration, tidal radius, or log-slope).  In \cite{minor2017} we 
showed that a robust mass can be defined in terms of the projected mass 
contained within the subhalo's perturbation radius $r_{\delta c}$. However, 
this robust mass estimator was tested using a simulated version of the ALMA 
lens SDP.81 for which the perturbation radius was very well-constrained; this 
is not always the case for many of our simulated lenses at HST resolutions with 
low-mass or low-concentration perturbations.

Thus, it is useful to test whether this mass estimator---the projected mass 
enclosed within the perturbation radius---is indeed unbiased for the subhalo 
fits shown in Figure \ref{fig:mbias}. To do this, we carry out the following 
procedure: 1) Calculate the perturbation radius $r_{\delta c}$ as a derived 
parameter during the nested sampling run. 2) After the run is finished, choose 
the median (50th percentile) inferred value of $r_{\delta c}$, which we call 
$r_{\delta c,fit}$  (one could also use the best-fit point for this). 3) For 
each point in the chain, calculate the subhalo's projected mass enclosed within 
the radius $r_{\delta c,fit}$, which we call $M_{2D,fit} \equiv 
M_{2D}(r_{\delta c,fit})$; this will be our mass estimator.  4) Plot the 
resulting posterior in $M_{2D,fit}$ and compare this to the actual projected 
mass within the \emph{same} radius, which we call $M_{2D,true}$. The latter 
point is important, since the true perturbation radius may differ from our 
inferred value.

The results of this procedure are shown in Figure \ref{fig:mpert}. The 
conventions are similar to Figure \ref{fig:mbias}, except now we are plotting 
the mass $M_{2D,fit}$ and comparing to $M_{2D,true}$ in each case, marked with 
an `x' for $z=0.2$ (or an asterisk for $z=0.5$). In nearly all cases $M_{2D}$ 
is recovered fairly well: for the $10^{10}M_\odot$ perturber 
(Figure \ref{fig:mpert_1e10}), $M_{2D,fit}$ differs from the true value by less than 
20\% in all cases; the same is true for the $10^9M_\odot$ and $10^8M_\odot$ 
perturbers except for the lowest-concentration case, for which the bias is 
26-27\% (albeit still within the error bars).  In all cases, the bias in the 
subhalo's total inferred mass is substantially larger.  For example, the 
$10^9M_\odot$ perturber with median concentration infers $M_{2D}$ with a bias 
of 8\%, whereas the total inferred mass $m_{200}$ is biased by 123\% (over a 
factor of 2). The numbers become starker for higher concentrations: for the 
cases with concentrations 2$\sigma$ above the median value, $M_{2D}$ is biased 
by a factor of 1.18, 0.90, and 0.99 for $\log m_{200}$ = 10, 9, and 8 
respectively; the corresponding bias factors in $m_{200}$ are 6.3, 3.6, and 
2.6. As expected, the bias in $M_{2D}$ was greatest in the cases for which the 
inferred perturbation radius $r_{\delta c}$ differed the most from its true 
value. However, even in cases where the true $r_{\delta c}$ lay outside the 
95\% probability region, the inferred $M_{2D,fit}$ was biased by less than 
20\%.

We conclude that the mass estimator from \cite{minor2017} is indeed robust, 
even at HST resolutions where the perturbation radius may be comparable to the 
pixel size, provided one follows the procedure outlined above. More generally, 
it has the additional advantage of providing physical insight into what is 
really being constrained during the lensing, as Section \ref{sec:mc_degeneracy} 
demonstrates. The perturbation radius can be calculated numerically via a standard root-finding algorithm; for details we refer the reader to Section 7.3 of \cite{minor2017}. In this way, $r_{\delta c}$ (and the corresponding subhalo mass enclosed within it) can be incorporated into existing lensing codes if desired and calculated as a derived parameter during lens modeling.

\begin{figure}
	\centering
	\includegraphics[height=0.8\hsize,width=0.9\hsize]{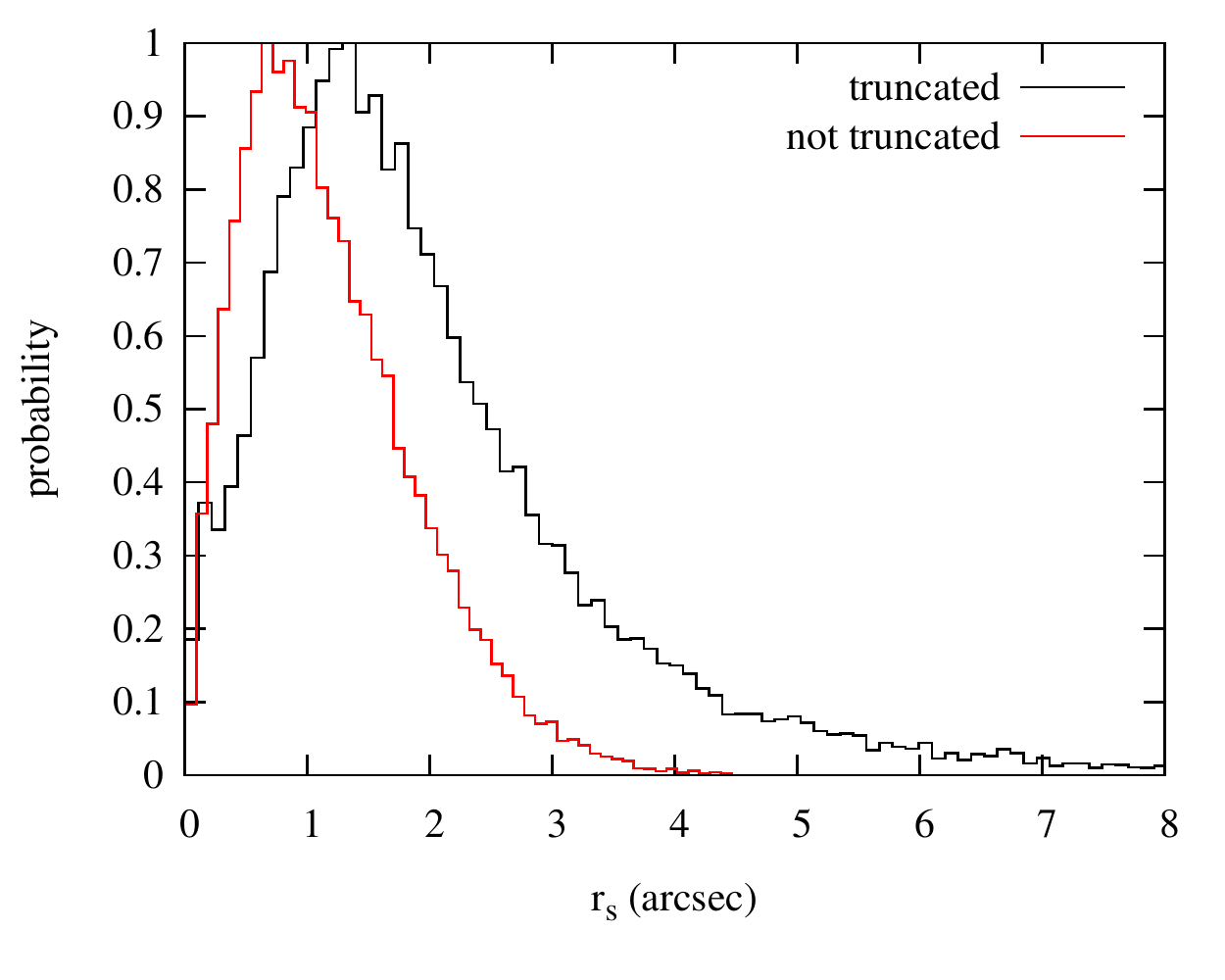}
\caption{Derived posteriors in the NFW scale radius $r_s$ of a $10^{10}M_\odot$ 
perturber at $z=0.2$. Black curve corresponds to a severely truncated subhalo 
with truncation radius $r_t \approx 0.29''$, which was modeled with $r_t$ fixed 
to this value. Red curve corresponds to a non-truncated $10^{10}M_\odot$ 
subhalo that generates a perturbation with the same scale $r_{\delta c} \approx 
0.058''$. Note that for the truncated subhalo, $r_s$ is allowed to be quite 
large (and hence, low $c_{200}$) because the perturbation is suppressed outside the 
truncation radius.  For most subhalos $r_t$ will be much greater than 
$r_{\delta c}$, so significant lower bounds on $c_{200}$ can be obtained.}
\label{fig:rs_posts}
\end{figure}

\section{Discussion: What really constrains the subhalo 
concentration?}\label{sec:c_limits}

In Section \ref{sec:mc_degeneracy} we have shown that the correlation between 
the inferred subhalo mass and subhalo concentration seen in sufficiently large 
perturbations (Figures \ref{fig:mcplot}, \ref{fig:mcplot_hires}) can be 
explained in terms of the mass enclosed within the perturbation radius being an 
approximately conserved quantity.  However this cannot be the \emph{only} 
physical constraint of interest, because it does not explain the upper/lower 
bounds on the inferred concentration.  Here, we discuss the origin of the 
constraints on the subhalo concentration.

\begin{figure*}
	\centering
	\subfigure[$c=2$ ~ $(m_{200}\approx2.4\times10^{12}M_\odot)$]
	{
		\includegraphics[height=0.40\hsize]{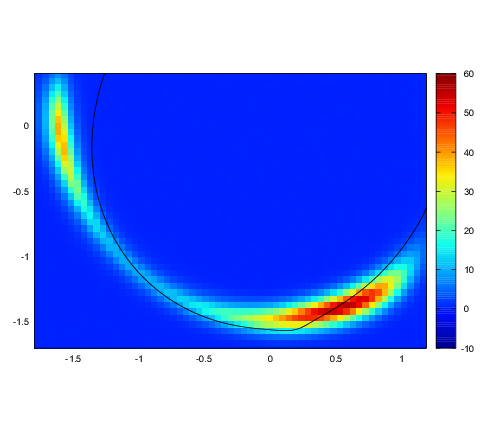}
		\label{fig:c2}
	}
	\subfigure[$c=10$ ~ $(m_{200}\approx5.0\times10^{10}M_\odot)$]
	{
		\includegraphics[height=0.40\hsize]{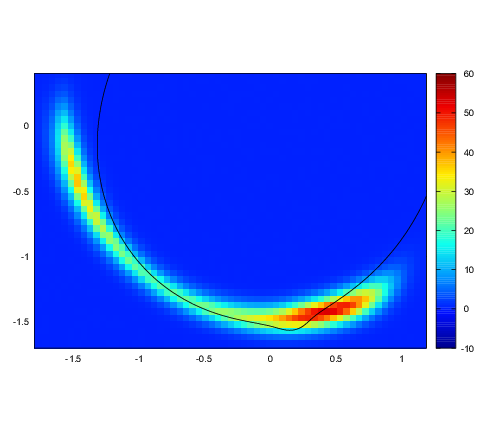}
		\label{fig:c10}
	}
	\subfigure[$c=40$ ~ $(m_{200}\approx3.0\times10^{9}M_\odot)$]
	{
		\includegraphics[height=0.40\hsize]{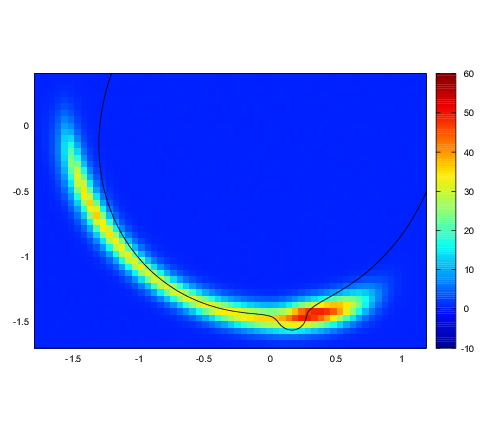}
		\label{fig:c40}
	}
	\subfigure[$c=160$ ~ $(m_{200}\approx8.2\times10^{8}M_\odot)$]
	{
		\includegraphics[height=0.40\hsize]{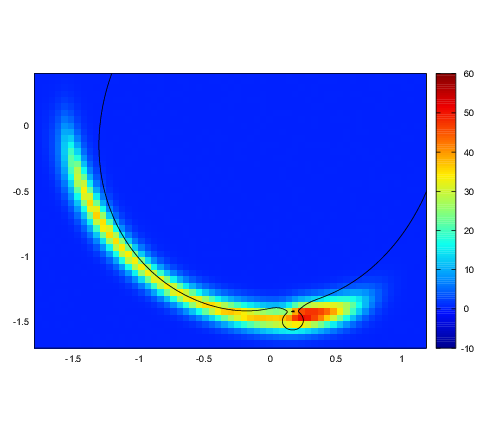}
		\label{fig:c160}
	}
	\caption{Comparison of the image perturbation produced by a subhalo with perturbation radius $r_{\delta c}=0.14''$, for four different concentrations: $c=2,10,40,160$. For each concentration, the subhalo mass $m_{200}$ is adjusted to produce the same $r_{\delta c}$. Note that for low concentrations ($c=2,10$), the images are perturbed far beyond the perturbation radius, whereas for high concentrations ($c=40,160$) only the immediate vicinity of the perturbation (up to a few times the perturbation radius) is affected. For reference, the perturbation size $r_{\delta c}=0.14''$ is the same as that produced by a $10^{10}M_\odot$ subhalo at the same location, with concentration $2\sigma$ above the median (blue contour in Figure \ref{fig:mcplot}).}
	\label{fig:ccomp}
\end{figure*}

In general, all of the subhalos that were detected in our mock data, no matter 
how small, produced a lower bound on the inferred concentration. This is the 
case even for $10^8M_\odot$ and low-concentration $10^9M_\odot$ subhalos for 
which the perturbation radius $r_{\delta c}$ is comparable to the pixel size, 
and hence is not well constrained. This indicates that the lower bound is not 
coming from the immediate neighborhood of the critical curve perturbation, but 
further out in the lens plane. Indeed, for sufficiently low concentrations, the 
images are visibly perturbed well beyond $r_{\delta c}$. This can be seen in 
Figure \ref{fig:ccomp}, where a subhalo with a perturbation radius $r_{\delta c}=0.14''$ is shown with different concentrations (this is the same $r_{\delta c}$ produced by a $10^{10}M_\odot$ subhalo with concentration $2\sigma$ above median, which corresponds to the blue contour in Figure \ref{fig:mcplot}). We plot the resulting images for $c=2,10,40,160$; note that the subhalo's $m_{200}$ is different in each figure, in order to produce the same $r_{\delta c}$ in each case. If the subhalo has too low of a 
concentration ($c=2,10$ in panels (a) and (b) respectively), the surface brightness far outside the perturbation radius is affected, as well as the overall size of the critical curve itself. To some extent, 
this can be compensated for by adjusting the primary lens parameters, in 
particular the Einstein radius and center coordinates, as well as the source 
galaxy parameters. However, for low enough concentrations, the subhalo mass 
must be very large to reproduce the critical curve  perturbation scale, and 
adjusting the primary lens/source parameters is not enough to achieve a good 
fit.

To demonstrate this, we can simulate and model a perturbation where the subhalo 
is severely truncated, forcing the perturbation to be more local. We generated 
a mock image using a truncated NFW profile for a $m_{200} = 10^{10}M_\odot$ 
(note this is the virial mass in the absence of truncation) with median 
concentration, and setting the truncation radius $r_t$ to be roughly 5 times 
the perturbation radius ($r_t \approx 0.29''$, $r_{\delta c} \approx 0.058''$).  
We then fit a subhalo with the truncation radius fixed to the correct value, 
thus forcing the perturbation to be small outside this region. In Figure 
\ref{fig:rs_posts} we plot the posterior probability in the NFW scale radius 
$r_s$, which is a derived parameter, where the black curve denotes the 
truncated fit. The red curve denotes the probability for an untruncated 
$10^{10}M_\odot$ subhalo, modeled without truncation, for comparison. \footnote{For the untruncated subhalo we chose the actual subhalo concentration to be 
slightly lower such that it matches the same perturbation radius ($r_{\delta c}$) as in 
the truncated subhalo. (Note that because the 3D profile is truncated, the 
projected density is reduced even near the subhalo center, reducing $r_{\delta 
c}$ compared to the same profile without truncation.)} The furthest image from 
the subhalo is approximately 2 arcseconds away, and this is roughly the limit 
on $r_s$ for the non-truncated subhalo. However the scale radius is allowed to 
be significantly larger (and thus, lower $c_{200}$) for the truncated subhalo, since 
the truncation prevents the perturbation from extending far beyond $r_t$. In 
reality, subhalos are unlikely to be tidally stripped far within $r_s$ without 
being disrupted altogether. In cases where $r_t \gtrsim r_s$, the truncation 
does not significantly affect the lower bound on $c_{200}$.

\begin{figure*}
	\centering
	\includegraphics[height=0.5\hsize,width=0.5\hsize]{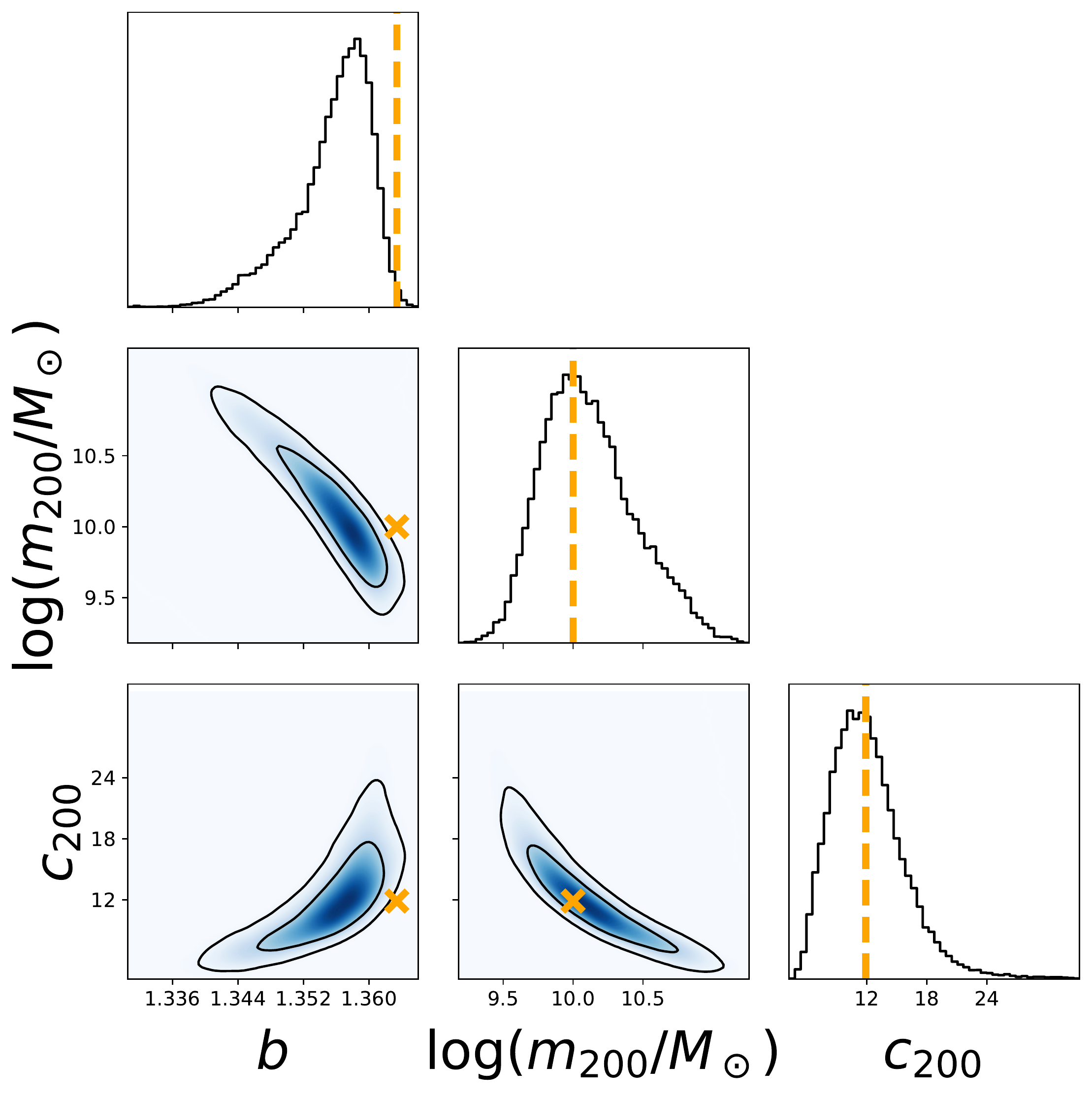}
\caption{Joint posteriors for a lens at $z=0.2$ perturbed by a subhalo at that is truncated at $r_{max} \approx 2.16r_s$, fit by an NFW profile without truncation. The parameters shown are the host galaxy Einstein radius $b$, subhalo log-mass $\log(m_{200})$, and subhalo concentration $c_{200}$. Note that the inferred subhalo mass and concentration are unbiased by the approximation of no truncation, while the host galaxy's Einstein radius comes out biased slightly low to offset the extra mass beyond the subhalo's actual tidal radius.}
\label{fig:mcb_posts}
\end{figure*}

In contrast to the lower limit on $c_{200}$, we find that the concentration on the 
high-end is only constrained if the perturbation radius is large enough that the 
\emph{shape} of the critical curve perturbation can be distinguished. This requires $r_{\delta c}$ to be at least as large as the PSF width (which in our simulations is roughly $0.05''$).  
Such is the case for all $10^{10}M_\odot$ subhalos in our mock data, as well as 
$10^9M_\odot$ subhalos with concentrations above the median $\Lambda$CDM 
value (for both $z=0.2$ and $z=0.5$). This can be seen in panels (c) and (d) of Figure \ref{fig:ccomp}, where the surface brightness in the vicinity of the perturbation (up to a few times $r_{\delta c}=0.14''$) can be distinguished between $c=40$ and $c=160$, allowing for the concentration to be constrained at the high end for this perturbation scale (blue contour in Figure \ref{fig:mcplot}). Such is also the case for the subhalo perturbing the lens SDSSJ0946+1006, whose concentration is higher than 3$\sigma$ above the expected median in CDM, generating $r_{\delta c}\sim0.3''$ as we show in a companion paper (Minor et al. 2020, in prep). However, for our mock data with a $10^9M_\odot$ subhalo with 
concentration at the median, we have $r_{\delta c} \approx 0.05''$ and 
there is no significant upper bound on the inferred concentration from the data, as can be seen in Figure \ref{fig:mcplot} (green contour).  
When the resolution is increased, as in Figure \ref{fig:mcplot_hires}, the 
perturbation is better resolved so that $c_{200}$ is constrained even for a 
low-concentration $10^9M_\odot$ subhalo, while a much stronger upper bound on 
the concentration is achievable for larger perturbations.

\section{Effect of tidal stripping on the inferred mass-concentration of subhalos}\label{sec:tidal_truncation}

We have seen that the concentration of $10^{10}M_\odot$ subhalos can be 
significantly constrained even at HST resolution (Figure \ref{fig:mcplot}), 
while lower mass subhalo concentrations can be constrained at higher 
resolutions (Figure \ref{fig:mcplot_hires}). However, our modeling up to now 
has ignored tidal stripping of subhalos. Since tidal truncation may mimic the 
effect of having a higher concentration to some extent, here we test whether 
the mass-concentration constraints are affected by tidal truncation.

$\Lambda$CDM simulations show that tidal stripping results in subhalos being 
truncated at or above $r_{max} \approx 2.16r_s$, while few subhalos are 
truncated below $r_{max}$ without severe tidal disruption occurring. Hence, we 
will simulate a truncated NFW subhalo with tidal radius $r_t = r_{max}$ and 
model it with an NFW profile to see if the resulting $m_{200}$ and $c_{200}$ are 
biased.  For truncated subhalos, we use the following ``smoothly truncated'' 
profile from \cite{baltz2009}:

\begin{equation}
\rho(r;r_s,r_t) = 
\frac{\rho_0}{\frac{r}{r_s}\left(1+\frac{r}{r_s}\right)^2\left(1+\left(\frac{r}{r_t}\right)^2\right)^2}
\label{eq:truncated_nfw}
\end{equation}
\vspace{5mm}
where $r_t$ is the tidal truncation radius. As our test case, we choose 
$m_{200} = 10^{10}M_\odot$ and $c_{200} = \bar c(M,z)$, solving for the appropriate 
$\rho_0$ and $r_s$ values for an NFW profile. We then truncate the profile 
using $r_t = r_{max} = 2.16r_s$ and generate simulated data using the same 
procedure as in \ref{sec:mockdata}.  The effect of truncation on the images is 
small within the perturbation radius, but quite noticeable outside it; the 
images near the critical curve are reduced slightly, while the farthest (and 
least magnified) image is displaced inwards roughly by one pixel-length.  In 
addition, the critical curve as a whole is slightly smaller.  
This begs the question whether subhalo inferences can be significantly biased by fitting a tidally stripped subhalo with an NFW profile without truncation.

To answer this, we fit this mock image assuming an NFW subhalo as in Section \ref{sec:mockdata}.  
Posteriors in $m_{200}$, $c_{200}$ and host galaxy Einstein radius $b$ are shown in Figure \ref{fig:mcb_posts}. Remarkably, there is 
no apparent bias in $m_{200}$ and $c_{200}$, despite completely ignoring the tidal 
truncation; however, the Einstein radius of the primary lens is biased low, in order to reduce the size of the critical curve in the absence of tidal truncation. This is accompanied by a simultaneous slight adjustment of the width and ellipticity of the source galaxy. One can see that in the limit of high concentration and low mass (approximating a truncated subhalo, in a rough sense), the 
Einstein radius approaches its actual value; in this limit however, the shape 
of the critical curve perturbation is noticeably altered in the vicinity of the 
perturbation radius (becoming ``sharper'' in appearance), degrading the fit.  
Hence, the best fit is achieved when the primary lens parameters are slightly 
adjusted, with the result that there is no discernible bias in the inferred 
subhalo concentration or mass parameters.

From this, we can conclude that tidal truncation is not expected to significantly bias the inferred subhalo mass for perturbers with CDM-like concentrations, at least at HST resolution. Nevertheless, it is possible that for high concentration perturbers, the effect of strong tidal truncation (up to $r_s$) is more local and not as easily mimicked by adjusting the primary lens model. For very strong perturbations, it is therefore advisable to include a tidal radius as a model parameter to reduce bias in the inferred subhalo parameters.

\section{Conclusions}\label{sec:conclusions}

We have demonstrated that the concentrations of dark matter subhalos perturbing a 
gravitationally lensed image are an important factor in the strength of the 
perturbation, allowing for joint constraints on the mass and concentration of 
subhalos. At HST-like resolutions, we show that the subhalo concentration can be constrained
for $\gtrsim 10^{10} M_\odot$ subhalos whose concentrations fall within the expected scatter in $\Lambda$CDM (Figure \ref{fig:mcplot}); constraints for lower mass subhalos may be possible if their concentrations are higher than the scatter expected in CDM. Looking to the future, we have shown that constraints on perturber concentration are achievable for 
$\gtrsim 10^8M_\odot$ subhalos at the $\sim0.01''$ resolution (Figure \ref{fig:mcplot_hires}) attainable by 
long-baseline interferometry and next-generation extremely large telescopes.  
This will make it possible to constrain the mass-concentration relation of dark matter halos on small scales from strong lensing perturbations. This approach is complementary to testing the expected mass function of $\Lambda$CDM at dwarf galaxy scales and provides a probe of the small-scale matter power spectrum in addition to dark matter physics.

Likewise, the concentration of perturbers plays an important role in the detectability of the perturbation. We have modeled a large number of mock lenses at HST resolution with subhalo perturbations involving a variety of masses and concentrations, in each case fitting a model with versus without a subhalo. For subhalos with masses of $10^8M_\odot$, the Bayesian evidence strongly favors a subhalo only if the subhalo is quite close to the critical curve \emph{and} if the concentration is at least $2\sigma$ above the expected median value in $\Lambda$CDM, indicating only the most concentrated $10^8M_\odot$ subhalos are likely to have detectable perturbations. For $10^9M_\odot$ subhalos, we find that perturbations are detectable over a broad range of positions (by the same criterion) only if their concentrations are at or above the median value in $\Lambda$CDM. In general, we conclude that perturbing subhalos of mass $<10^9M_\odot$ may not be detected unless they have concentrations above the median expected value for $\Lambda$CDM.

We have also shown that if scatter in the mass-concentration relation is unaccounted for during lens modeling, the inferred subhalo masses can be biased 
biased by a factor of 3(6) for subhalos of mass $10^9 M_\odot$($10^{10} 
M_\odot$) (Figure \ref{fig:mbias}). This bias can be eliminated in one of two ways: 1) varying both the 
concentration and mass as free parameters; 2) instead of inferring the total mass, inferring the projected mass  
within the subhalo's perturbation radius, defined by its distance to the 
critical curve of the lens at the point of maximum perturbation. The latter can be determined much more robustly than the total subhalo mass, with little degeneracy with concentration or density slope, as demonstrated in Figure \ref{fig:mpert_vs_logslope}. In practice, both 
strategies can be combined, providing constraints on both $c_{200}$ and $m_{200}$ for perturbing halos under the assumption of an NFW profile, while also providing a robust mass measurement that holds 
even if the assumed density profile is incorrect (e.g. if a dark matter larger
core is present). We use this approach to model the substructure in the lens SDSSJ0946+1006 in a companion paper (Minor et al. 2020, in prep).

Finally, we have shown that for HST resolutions, even strong tidal truncation has a negligible bias on the inferred subhalo parameters for high-mass subhalos with CDM-like concentrations (Figure \ref{fig:mcb_posts}). We caution, however, that the bias may be more significant for highly concentrated perturbers or at higher resolutions; in such cases, the safest approach may be to include tidal truncation as a model parameter for the subhalo.

As next-generation sky surveys and ground-based extremely large
telescopes come online, the ability to measure subhalo concentrations will provide a critical test of the cold dark matter paradigm. The resulting constraints, perhaps in concert with other probes, may provide valuable insights into dark matter physics and the small-scale matter power spectrum.

\section*{Acknowledgements}

We thank Anna Nierenberg for useful discussions during the course of the 
project. QM was supported by NSF grant AST-1615306 and MK by NSF PHY-1915005.

We gratefully acknowledge a grant of computer time from XSEDE allocation 
TG-AST130007.


\bibliography{subhalo2}



\end{document}